\documentclass{article}
\usepackage{graphicx} 
\usepackage{color}
\usepackage{subcaption}
\usepackage{amsmath}
\usepackage[style=numeric-comp,sorting=none,minnames=10,maxnames=99,maxbibnames=99,giveninits=true]{biblatex}
\usepackage{authblk}
\usepackage{cleveref}
\usepackage{mathrsfs}
\usepackage[ruled,linesnumbered,noline]{algorithm2e}
\usepackage[%
    a4paper,
    left=4.cm,
    right=4.cm,
    top=3cm,
    bottom=3cm,
    headheight=15pt,
    footskip=1cm,
    includehead
]{geometry}

\title{An Improved Boris Algorithm for {\color{blue}Charge Particle Orbit} in Tokamak Plasmas}

\author[1,2]{Jian Wang\thanks{Corresponding Author. Email address: wang.jian@ipp.ac.cn}}
\author[1]{Xiaodong Zhang}
\author[1]{Lei Ye}
\author[1]{Xingyuan Xu}

\affil[1]{Institute of Plasma Physics, Hefei Institutes of Physical Science, Chinese Academy of Science, Hefei 230031, China}

\affil[2]{University of Science and Technology of China, Hefei 230026, China}

\date{}

\begin{document}

\maketitle

\begin{abstract}
     An improved Boris algorithm for simulating the motion of charged particles in electromagnetic fields has been developed. This enhancement addresses the issue of inaccurate fast-scale cyclotron phase calculations present in the original Boris algorithm, while preserving its advantage in simulating slow-scale guiding center motion. As a result, it strikes a balance between low and high-frequency dynamics, overcoming the limitations of traditional second-order volume-preserving algorithms (VPAs) which are constrained to a single characteristic frequency. Test particle simulations indicate that, in most cases, the improved Boris algorithm achieves significantly higher accuracy than conventional VPAs when simulating cases involving various frequencies of electric field within a typical Tokamak magnetic field, highlighting its superior efficacy in handling problems across a wide range of characteristic frequencies. 
     
\end{abstract}

\section{Introduction}
~~~~Accurate numerical calculations of charged particle dynamics within electromagnetic fields are essential for plasma simulations. A variety of numerical integration methods have been developed, with the fourth-order Runge-Kutta method {\color{blue}(RK4)} and the Boris algorithm \cite{Boris,Computer_Simulations_Using_Particles,Plasma_Physics_via_Computer_Simulation} being particularly representative. As a widely employed numerical method for solving differential equations, the RK4 method, despite its high accuracy of fourth-order, suffers from rapid error accumulation over extended computational durations,  ultimately compromising its effectiveness in long-term simulations \cite{QinPoP2013}. In contrast, while the Boris algorithm offers relatively lower precision with second-order accuracy, it exhibits exceptional stability across a broad range of temporal scales \cite{ParkerJCP1991-Boris,Stoltz2002-Boris,Penn2003-Boris}, making it highly compatible with the multi-scale nature of plasma physics. Modifications to the original Boris algorithm have been proposed to address specific scenarios, such as strong magnetic field environments \cite{Hairer2020-1,Hairer2020-2,Hairer2022}, relativistic dynamics \cite{DECYK2023108559} and so on, all of which have consistently produced reliable results. 

The success of the Boris algorithm can be attributed to its ability to preserve phase space volume \cite{QinPoP2013}. Algorithms possessing this property are referred to as volume-preserving algorithms {\color{blue}(VPAs)}, and can be conveniently represented using splitting techniques of Lie algebra \cite{ZhangCCP2015-LieAlgebra,PhysRevE.92.063310} or matrix notations \cite{wang2025}. In this sense, the Boris algorithm is classified as a second-order VPA. Another second-order VPA, denoted as $G_h^2$, was first proposed in \cite{HeJCP2015-VPA}. It is {\color{blue}obtained} by modifying the magnetic-field-induced rotation angle of the velocity variable in the Boris algorithm, thereby preserving phase space volume while exhibiting distinct numerical characteristics and performance compared to the original Boris algorithm \cite{HeJCP2015-VPA,HeJCP2016-Higher_order_VPA}. Furthermore, higher-order VPAs have been developed and shown to deliver satisfactory results in simulating both non-relativistic and relativistic dynamics \cite{HeJCP2016-Higher_order_VPA,HePoP2016-Relativistic_VPA,WangPoP2016-Runaway_Electrons}. 

In Tokamak plasmas, the motion of charged particles is characterized by two distinct scales: the slow, low-frequency transit/bounce motion of the guiding center, and the fast, high-frequency cyclotron motion. The theoretical phase stability analysis presented in \cite{wang2025} indicates that, the Boris algorithm and $G_h^2$ are the most efficient {\color{blue}Boris-like} second-order VPA for capturing these two typical scales, respectively. This conclusion is also strongly corroborated by numerical experiments, underscoring the necessity of selecting an appropriate VPA according to the dominant frequency scale of the problem at hand. 

Nevertheless, conventional VPAs are intrinsically constrained by a single characteristic frequency. The Boris algorithm, while optimal for capturing slow-scale guiding center motion, exhibits poor convergence in resolving fast-scale cyclotron phases. Conversely, $G_h^2$ performs well in simulating cyclotron motion but introduces cumulative drift in the guiding center trajectory over time. In this paper, we propose a novel algorithm that integrates the strengths of both methods, effectively combining their respective advantages in handling low- and high-frequency dynamics. Numerical experiments demonstrate that the proposed algorithm achieves significantly higher accuracy and efficiency compared to conventional VPAs, highlighting its strong potential and broad applicability in the simulation of charged particle dynamics. 

This paper is organized as follows. Section II provides a detailed formulation of the improved Boris algorithm. In Section III, the numerical accuracy and efficiency of the proposed method are evaluated and compared with conventional VPAs through simulations of charged particle dynamics in a typical Tokamak toroidal magnetic field with varying electric fields. Finally, Section IV concludes the paper.

\section{Construction of the Improved Boris Algorithm}
~~~~~~This section is dedicated to the construction of the improved Boris Algorithm. The formulation is articulated through matrix notation, following the framework established in \cite{wang2025}. 

Before introducing the detailed construction process, we briefly review the generalized form of {\color{blue}Boris-like} second-order volume-preserving algorithms. The motion of charged particles in an electromagnetic field $\vec E=(E^x,E^y,E^z)^T$ and $\vec B=(B^x,B^y,B^z)^T$ is governed by the Lorentz-Newton equation

$$m \frac{d\vec v}{dt}=q(\vec v \times \vec B+\vec E) \eqno{(1.a)} $$
$$\frac{d\vec r}{dt}=\vec v \eqno{(1.b)}$$
with $m$ the mass, $q$ the electric charge, and $\vec r=(x,y,z)^T$, $\vec v=(v^x,v^y,v^z)^T$ the position and velocity of the charged particle under Cartesian coordinates. 

To facilitate the discussion, the magnetic field $\vec B$, electric field $\vec E$, velocity variable $\vec v$, position variable $\vec r$ time variable $t$ are normalized by basic quantities 
$$
B_{ref} = B_0,v_{ref} = v_0
\eqno{(2.a)} 
$$
with $B_0$ the magnetic field strength on the magnetic axis, $v_0$ the initial velocity magnitude of the particle. And derived quantities are given by
$$
E_{ref}=B_{ref} v_{ref}=B_0v_0,t_{ref}=\frac{m}{qB_{ref}}=\frac{m}{qB_0},r_{ref}=v_{ref} t_{ref} = \frac{mv_0}{qB_0}
\eqno{(2.b)} 
$$
Replacing $\vec B, \vec E,\vec v,\vec r$ and $t$  by $\frac{\vec B}{B_{ref}},\frac{\vec E}{E_{ref}},\frac{\vec v}{v_{ref}},\frac{\vec r}{r_{ref}}$ and $\frac{t}{t_{ref}}$ in Equations (1) yields
$$\frac{d\vec v}{dt}=\vec v \times \vec B+\vec E \eqno{(3.a)} $$
$$\frac{d\vec r}{dt}=\vec v \eqno{(3.b)}$$
Throughout the discourse in this section, we shall persistently utilize the above normalized form. 

The generalized form of {\color{blue}Boris-like} second-order volume-preserving algorithm in matrix notation, as given in \cite{wang2025}, is expressed as follows
$$
\vec v_{k+1}=R_k \vec v_k + \frac{\Delta t}{2}(I+R_k)\vec E_k
\eqno{(4.a)}
$$
$$
\vec r_{k+1}=\vec r_k+\Delta t \cdot \vec v_{k+1}=\vec r_k +{\Delta t}\cdot R_k \vec v_k+\frac{\Delta t^2}{2}(I+R_k){\vec E_k}
\eqno{(4.b)}
$$
Here, $\Delta t$ denotes the fixed time step size. Let $t_j=j\cdot \Delta t$ denotes time grid for arbitrary $j$, then $\vec v_k=\vec v(t_k), \vec r_k=\vec r(t_{k+\frac{1}{2}}), \vec B_k=\vec B(\vec r_k,t_{k+\frac{1}{2}}), \vec E_k=\vec E(\vec r_k,t_{k+\frac{1}{2}})$ denotes variables at the  $k-th$  time step. And $R_k$ is the rotation matrix, defined as follows
$$
R_k=P_k \Lambda_k P_k^*
\eqno{(4.c)}
$$
$$
P_k=\frac{1}{B_k}
\begin{bmatrix}
 B_k^x & \frac{-B_k^x B_k^y - B_k^z B_k i}{\sqrt{2((B_k^x)^2+(B_k^z)^2)}} & \frac{B_k^x B_k^y - B_k^z B_k i}{\sqrt{2((B_k^x)^2+(B_k^z)^2)}}\\
 B_k^y & \frac{\sqrt{(B_k^x)^2+(B_k^z)^2}}{\sqrt{2}} & -\frac{\sqrt{(B_k^x)^2+(B_k^z)^2}}{\sqrt{2}}\\
 B_k^z & \frac{-B_k^y B_k^z + B_k^x B_k i}{\sqrt{2((B_k^x)^2+(B_k^z)^2)}} & \frac{B_k^y B_k^z + B_k^x B_k i}{\sqrt{2((B_k^x)^2+(B_k^z)^2)}}\\
\end{bmatrix}
\eqno{(4.d)}
$$
$$
\Lambda_k=\text{diag}(1,\text{exp}(\theta_k \cdot i),\text{exp}(-\theta_k \cdot i))
\eqno{(4.e)}
$$
where $P_k$ is a unitary matrix, and $P_k^*$ represents its conjugate transpose. $B_k=\sqrt{(B_k^x)^2+(B_k^y)^2+(B_k^z)^2}$ denotes the magnetic field strength, and $\theta_k$ is the magnetic-field-induced rotation angle of the velocity variable $\vec v_k$, which needs to satisfy the condition of consistency
$$
\lim_{\Delta  t\to 0} \frac{\theta_k}{{B_k}\cdot \Delta t}=1 
\eqno{(4.f)}
$$
In other words, by treating $\theta_k$ as a function of the time step size $\Delta t$ while simultaneously ensuring the condition of consistency (4.f), one can derive a series of second-order volume-preserving algorithms. The well-known Boris algorithm corresponds exactly to the case where $\theta_k=2\text{arctan}(\frac{1}{2} {B_k} \cdot \Delta t)$ . An alternative valid formulation, obtained by simply taking $\theta_k=B_k \cdot \Delta t$, leads to another volume-preserving scheme known as $G_h^2$, which was initially proposed in \cite{HeJCP2015-VPA} using Lie algebra techniques and represented via matrix exponentials.

Now we are in the position to construct the improved algorithm. Theoretical phase stability analysis in \cite{wang2025} indicates that the two aforementioned algorithms are optimally suited for low- and high-frequency dynamics, respectively, within this class of {\color{blue}Boris-like} second-order volume-preserving algorithms defined by Equations (4). Specifically, the Boris algorithm proves to be the most effective scheme for capturing slow-scale guiding center motion (i.e., low-frequency dynamics), although it suffers from significant errors when resolving fast-scale cyclotron motion. In contrast, $G_h^2$ excels in modeling high-frequency cyclotron motion but introduces noticeable deviations in guiding center trajectories during long-term simulations. To overcome the inherent limitation of existing algorithms being constrained to a single characteristic frequency, we propose a hybrid approach that combines the strengths of both methods. Such an integration enables accurate treatment of both low- and high-frequency dynamics, thereby enhancing overall numerical accuracy while reducing computational cost. The Boris algorithm is selected as the foundational scheme due to its superior performance in slow-scale dynamics and its robustness in long-term computations. This choice also motivates naming the new approach the 'Improved Boris Algorithm'. Simultaneously, $G_h^2$ is incorporated to provide reliable information on cyclotron motions and velocity variables, which must be periodically recalibrated to prevent trajectory deviation. 

Building on the above ideas, we now present the specific implementation of the improved algorithm. Let $\vec r_{1,k}=\vec r_{Boris}(t_{k+\frac{1}{2}})$, $\vec v_{1,k}=\vec v_{Boris}(t_k)$, $\vec r_{2,k}=\vec r_{G_h^2}(t_{k+\frac{1}{2}})$, and $\vec v_{2,k}=\vec v_{G_h^2}(t_k)$ denote the variables calculated by the Boris algorithm and $G_h^2$, respectively, while $\vec r_k=\vec r(t_{k+\frac{1}{2}})$ and $\vec v_k=\vec v(t_k)$ represents the numerical results of the new algorithm at the $k-th$ time step. It appears that we can directly derive
$$
\vec v_{k}=\vec v_{2,k}
\eqno{(5.a)}
$$
$$
\vec r_{k}=\vec r_{1,k}^{GC}+\vec r_{2,k}^{C}
\eqno{(5.b)}
$$
Here, the superscript 'GC' and 'C' represent the guiding center and the cyclotron trajectory, respectively, as given by the following equations (cf. \cite{Hairer2020-1}, \cite{LiuPoP2011})
$$
\vec r^C=-\frac{\vec v \times \vec B(\vec r ,t)}{B^2(\vec r ,t)}
\eqno{(5.c)}
$$
$$
\vec r^{GC}=\vec r-\vec r^C=\vec r+\frac{\vec v \times \vec B(\vec r ,t)}{B^2(\vec r ,t)}
\eqno{(5.d)}
$$
Equation (5.b) implies that the position variable of the new algorithm is obtained by combining the guiding center trajectory (low-frequency component) from the Boris algorithm and the cyclotron trajectory (high-frequency component) from $G_h^2$, achieving the objective of leveraging the strengths of both algorithms. Meanwhile, in typical scenarios of Tokamak plasmas, the cyclotron velocity of the particles is significantly greater than the drift velocity. Thus, it is reasonable to adopt the velocity directly from $G_h^2$, as reflected in Equation (5.a).  

However, the position and velocity variables in both VPAs differ by half a time step to ensure {\color{blue}volume-preserving property}, resulting in $O(\Delta t)$ errors when directly using Equations (5.c) and (5.d) to compute the guiding center and cyclotron trajectories. To address this issue, we consider the central difference for Equation (3.a)
$$
\frac{\vec v_{i,k+1}-\vec v_{i,k}}{\Delta t}=\vec v_{i,k+\frac{1}{2}}\times \vec B_{i,k}+\vec E_{i,k}+O(\Delta t ^2)
\eqno{(6)}
$$
with the subscript $i=1,2$ refers to the results from the Boris algorithm and $G_h^2$, respectively. Substituting Equation (6) into Equations (5.c) and (5.d) yields
$$
\vec r_{i,k}^C=\frac{1}{B^2_{i,k}} \left[ \vec E_{i,k}-\frac{\vec v_{i,k+1}-\vec v_{i,k}}{\Delta t}\right]+O(\Delta t^2)
\eqno{(7.a)}
$$
$$
\vec r_{i,k}^{GC}=\vec r_{i,k}-\vec r_{i,k}^C=\vec r_{i,k}-\frac{1}{B^2_{i,k}} \left[ \vec E_{i,k}-\frac{\vec v_{i,k+1}-\vec v_{i,k+1}}{\Delta t}\right]+O(\Delta t^2)
\eqno{(7.b)}
$$
Through this correction, both $\vec{r}_{i,k}^C$ and $\vec{r}_{i,k}^{GC}$ are evaluated at $t_{k+\frac{1}{2}}$, consistent with the temporal location of $\vec{r}_{i,k}$. These expressions retain second-order accuracy, ensuring that no additional numerical errors are introduced into the improved algorithm. 

As previously mentioned, it is necessary to periodically reset the results of $G_h^2$ to prevent cumulative trajectory deviations. In contrast, the Boris algorithm maintains consistent accuracy over long-term simulations without deviation. Furthermore, our numerical tests reveal that the Boris algorithm is extremely sensitive to initial conditions. Namely, any form of recalibration—which is equivalent to essentially resetting its initial conditions—can result in significant degradation of accuracy.  On the other hand, $G_h^2$ exhibits strong robustness against such sensitivity. Consequently, recalibration is applied exclusively to $G_h^2$. For simplicity, the time interval between two successive recalibrations, namely the recalibration period $\Delta T$, is treated as a fixed constant. Therefore, recablirations are performed at the $n-th$ time step where $n\cdot \Delta t=c\cdot \Delta T$ with $c$ being an arbitrary positive integer.     

In conclusion, the specific derivation process of the Improved Boris Algorithm is outlined below (Algorithm \ref{algorithm}). {\color{blue}The corresponding explicit update expressions for the new algorithm at the $(k+1)-th$ time step are as follows
$$
\vec v_{1,k+1}=R_{1,k}\vec v_{1,k}+\frac{\Delta t}{2}(I+R_{1,k})\vec E_{1,k}
\eqno{(8.a)}
$$
$$
\vec r_{1,k+1}=\vec r_{1,k}+\vec v_{1,k+1}\cdot \Delta t=\vec r_{1,k}+\Delta t \cdot R_{1,k}\vec v_{1,k}+\frac{\Delta t^2}{2}(I+R_{1,k})\vec E_{1,k}
\eqno{(8.b)}
$$
$$
\vec v_{2,k+1}=R_{2,k}\vec v_{2,k}+\frac{\Delta t}{2}(I+R_{2,k})\vec E_{2,k}
\eqno{(8.c)}
$$
$$
\vec r_{2,k+1}=\vec r_{2,k}+\vec v_{2,k+1}\cdot \Delta t=\vec r_{2,k}+\Delta t \cdot R_{2,k}\vec v_{2,k}+\frac{\Delta t^2}{2}(I+R_{2,k})\vec E_{2,k}
\eqno{(8.d)}
$$
$$
\vec v_{k+1}=\vec v(t_{k+1})=\vec v_{2,k+1}=R_{2,k}\vec v_{2,k}+\frac{\Delta t}{2}(I+R_{2,k})\vec E_{2,k}
\eqno{(8.e)}
$$
$$
\vec r_{k}=\vec r(t_{k+\frac{1}{2}})=\vec r_{1,k}-\frac{1}{B_{1,k}^2} \left[\vec E_{1,k}-\frac{\vec v_{1,k+1}-\vec v_{1,k}}{\Delta t} \right]+\frac{1}{B_{2,k}^2}\left[\vec E_{2,k}-\frac{\vec v_{2,k+1}-\vec v_{2,k}}{\Delta t} \right]
\eqno{(8.f)}
$$
Here, the subscript $(1,k)$ refers to quantities computed using the Boris algorithm, $(2,k)$ corresponds to those from $G_h^2$, and $(k)$ the outputs of the improved Boris algorithm, as previously defined.

Their initial conditions are set as
$$
\vec v_{1,0}=\vec v_{2,0}=\vec v_0=\vec v(0)
\eqno{(9.a)}
$$
$$
\vec r_{1,0}=\vec r_{2,0}=\vec r_0=\vec r(0)+\frac{\Delta t}{2}\vec v(0)
\eqno{(9.a)}
$$

For the recalibration step, an additional adjustment to the numerical results of $G_h^2$ ($\vec r_{2,k+1}$) is also required
$$
\vec r_{2,k+1}=\vec r_k+\Delta t \cdot \vec v_{k+1}
\eqno{(10)}
$$
}
\begin{algorithm}
\caption{Improved Boris Algorithm}\label{algorithm}
\KwData{initial conditions of position $\vec r_0$ and velocity $\vec v_0$, total calculation time $T$, fixed recalibration period $\Delta T$, fixed time step size $\Delta t$.}
\KwResult{numerical solutions of position $\vec r(T)$ and velocity $\vec v(T)$.}
~~~~[Comment: For simplicity, it is assumed that the total calculation time $T$ is an integer multiple of the time step size $\Delta t$. Otherwise, an additional adjustment of the time step size is required at the end of the algorithm.]

$t\leftarrow 0$, $k \leftarrow 0$, $t_1\leftarrow 0$\;
~~~~[Comment: Here, $t$ represents the current calculation time, $k$ represents the current time step and $t_1$ serves as the timer for recalibration.]

$\vec v_{1,0} \leftarrow \vec v_0$, $\vec v_{2,0} \leftarrow \vec v_0$\;
~~~~[Comment: Initialization of velocity variables for the Boris algorithm and $G_h^2$.]

$\vec r_{1,0} \leftarrow \vec r_0 + \frac{1}{2} \vec v_0 \cdot \Delta t$, $\vec r_{2,0} \leftarrow \vec r_0 + \frac{1}{2} \vec v_0 \cdot \Delta t$\;
~~~~[Comment: Initialization of position variables for both VPAs. The position variables are advanced by half a time step relative to the velocity variables to ensure second-order accuracy.]

\While{$t < T$}
{
Calculate $\vec v_{1,k+1}$ by Equation (8.a)\;
Calculate $\vec v_{2,k+1}$ by Equation (8.c)\;
$\vec v_{k+1} \leftarrow \vec v_{2,k+1}$\;
~~~~[Comment: Update the velocity variables for both VPAs. The result of the new algorithm($\vec v_{k+1}$) is equivalent to that of $G_h^2$($\vec v_{2,k+1}$).]

Calculate $\vec r_{1,k}^{GC}$ by Equation (7.b)\;
Calculate $\vec r_{2,k}^{C}$ by Equation (7.a)\;
$\vec r_k \leftarrow \vec r_{1,k}^{GC}+r_{2,k}^{C}$\;
~~~~[Comment: The position variable of the new algorithm($\vec r_k$) is obtained by combining the guiding center trajectory from the Boris algorithm($\vec r_{1,k}^{GC}$) and the cyclotron trajectory from $G_h^2$($\vec r_{2,k}^{C}$).]

$\vec r_{1,k+1} \leftarrow \vec r_{1,k} + \vec v_{1,k+1} \cdot \Delta t$\;
$\vec r_{2,k+1} \leftarrow \vec r_{2,k} + \vec v_{1,k+1} \cdot \Delta t$\;
~~~~[Comment: Update the position variables for both VPAs.]

$t_1 \leftarrow t_1+\Delta t$\;
\If{$t_1 \geq \Delta T$}
{
$\vec r_{2,k+1} \leftarrow \vec r_{k} + \vec v_{k+1} \cdot \Delta t$\;
$t_1 \leftarrow 0$\;
}
~~~~[Comment: Recalibrate the result of $G_h^2$ by the result of the new algorithm.]

$t \leftarrow t+\Delta t$\;
$k \leftarrow k+1$\;
}
$\vec r_{k+1} \leftarrow \vec r_k + \frac{1}{2} \vec v_{k+1} \cdot \Delta t$\;
$\vec r(T) \leftarrow \vec r_{k+1}, \vec v(T) \leftarrow \vec v_{k+1}$\;
\end{algorithm}

{\color{blue} From Equations (8), it is evident that the numerical results of the new algorithm at each step are derived through a linear combination of the outputs from two volume-preserving algorithms evaluated at the same time step. In other words, there is no explicit mapping between the phase space coordinates at the $k-th$ time step $(\vec v_k,\vec r_{k-1})$ and those at the $(k+1)-th$ time step $(\vec v_{k+1},\vec r_{k})$, which makes it considerably challenging to theoretically verify its volume-preserving property from the original definitions. 

Nevertheless, in the absence of the electric field, the algorithm still exhibits a key feature of traditional VPAs, namely, exact conservation of kinetic energy through preservation of velocity magnitude. From Equation (8.a) we obtain 
$$
\vec v_{k+1}=R_{2,k}\vec v_{2,k}=R_{2,k}\vec v_{k}
\eqno{(11)}
$$
$R_{2,k}$ is orthogonal for arbitrary $k$, as it can be expressed as the product of three unitary matrices (Equation (4.c)), and it is also real-valued. Hence, the velocity magnitude remains constant across all time steps, namely $v_k=v_{k-1}=...=v_0$ in this case. 

Besides, as a composite of volume-preserving algorithms, its long-term numerical behavior is expected to closely resemble that of its constituents, especially with the conventional Boris algorithm. Therefore, its stability over extended simulation durations is effectively inherited from the intrinsic stability of the underlying volume-preserving algorithms.
}

\begin{figure}[tbp]
\centering
\includegraphics[width=1.0\textwidth]{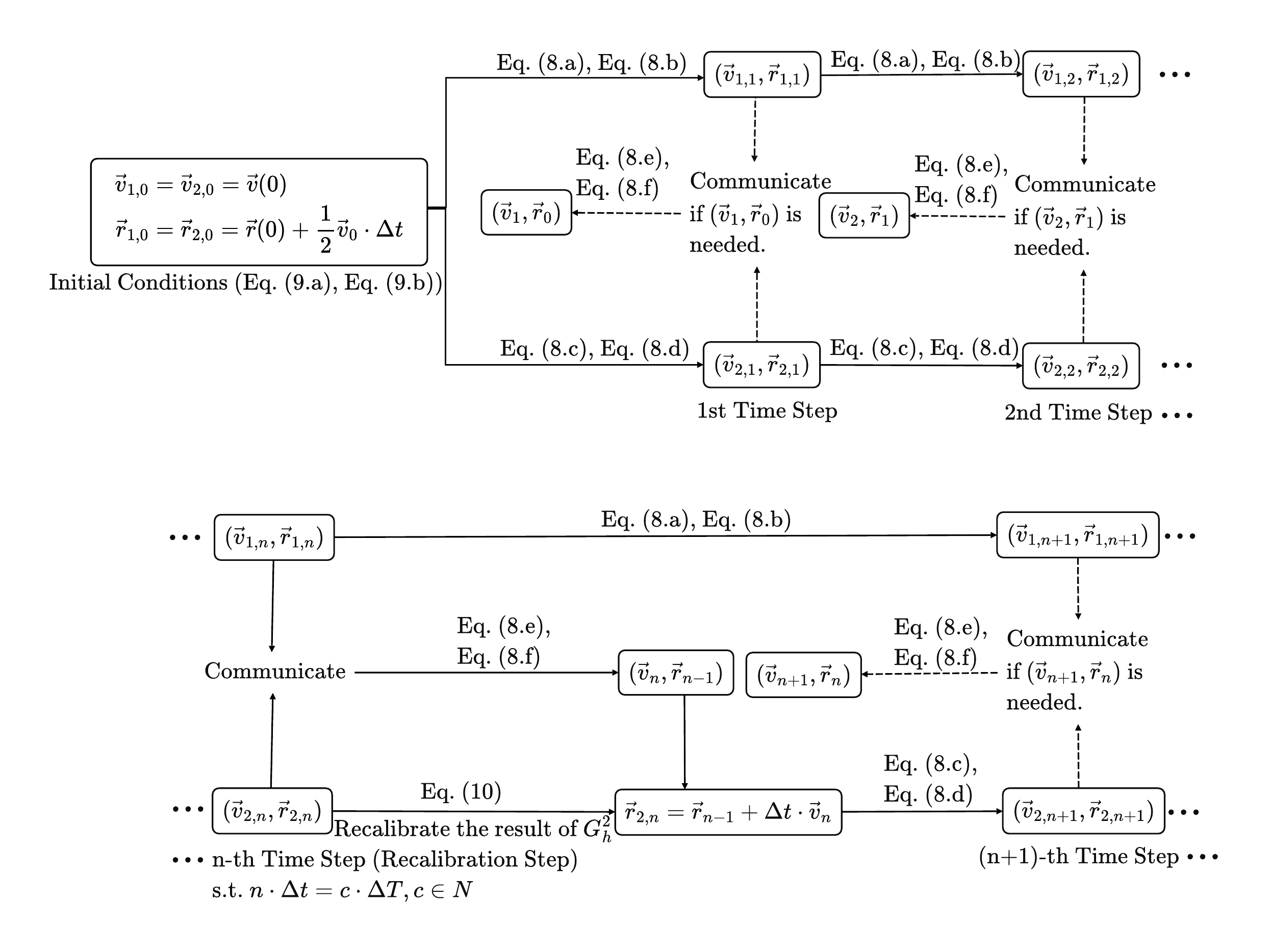}
\caption{Paralellized plan for the improved Boris algorithm.}
\label{fig 1-1}
\end{figure}

From the process described, it is evident that the theoretical computational cost of the Improved Boris Algorithm is the sum of the costs of the Boris algorithm and $G_h^2$. At first glance, this may suggest that the algorithm is not truly "improved" in terms of computational efficiency. {\color{blue}However, the computational procedures of the Boris algorithm and $G_h^2$ are inherently independent and can be executed in parallel, as illustrated in Figure~\ref{fig 1-1}. Clearly, data communication and synchronization between the two parallel tasks are required only at specific time steps—namely, those where the results of the improved algorithm are needed, and during the recalibration steps. This parallelization would reduce the computational time of the new algorithm to that of $G_h^2$ (Since $G_h^2$ involves calculations of trigonometric functions, while the Boris algorithm does not, making $G_h^2$ slightly more time-consuming in comparison)}. Besides, as will be demonstrated in the numerical experiments conducted in the next section, the new algorithm achieves significantly higher accuracy than both VPAs under identical conditions in the vast majority of cases.

\section{Numerical Experiments}
~~~~In this section, numerical experiments are performed to evaluate the accuracy and efficiency of the Improved Boris Algorithm {\color{blue}(non-parallelized)}, in comparison with the conventional Boris algorithm and $G_h^2$. All three algorithms are implemented in C++, and all numerical calculations are carried out in double precision with a single core of the 3.0 GHz Intel i9-13900k processor using the GCC compiler.  

In the following analysis, physical quantities are no longer normalized. Instead, the time variables are qualified by the gyro-frequency of an ion in a fixed magnetic field {\color{blue}$B_0=1\ T$}
$$
\color{blue} \omega_{c0}=\frac{eB_0}{m_i}=9.57\times 10^7\ s^{-1}
\eqno{(12)}
$$

Consider the motion of a single ion in a toroidal magnetic field with magnetic field strength on the magnetic axis {\color{blue}$B_{axis}=2\ T$}, major radius {\color{blue}$R_0=1.67\ m$}, minor radius {\color{blue}$a=0.6\ m$}, and the safety factor 
$$
q=2.52(\frac{r}{a})^2-0.16(\frac{r}{a})+0.86
\eqno{(13)}
$$
with $r=\sqrt{(\sqrt{x^2+y^2}-R_0)^2+z^2}$. These parameters can be referenced in \cite{GorlerPoP2016-GYRO}. The magnetic field  in toroidal coordinates $(r,\theta,\phi)$ is expressed as $\vec B=B_{\theta} \vec e_{\theta}+B_{\phi} \vec e_{\phi}$ with
$$
B_{\phi}=\frac{B_{axis}R_0}{R_0+r \text{cos}\theta},B_{\theta}=\frac{rB_{\phi}}{qR_0}
\eqno{(14)}
$$
To apply the algorithms, we transform the toroidal magnetic field $\vec B$ into the Cartesian coordinates $(x,y,z)$ which is
$$
B_x=-B_{\phi} \text{sin}\phi-B_{\theta} \text{sin}\theta \text{cos}\phi=-\frac{B_{axis}R_0 y}{x^2+y^2}-\frac{B_{axis}xz}{q(x^2+y^2)}
\eqno{(15.a)}
$$
$$
B_y=B_{\phi} \text{cos}\phi-B_{\theta} \text{sin}\theta \text{sin}\phi=\frac{B_{axis}R_0 x}{x^2+y^2}-\frac{B_{axis}yz}{q(x^2+y^2)}
\eqno{(15.b)}
$$
$$
B_z=B_{\theta} \text{cos}\theta=\frac{B_{axis}(\sqrt{x^2+y^2}-R_0)}{q\sqrt{x^2+y^2}}
\eqno{(15.c)}
$$
We will consider the motion of particles in electric fields of various strengths and frequencies within the aforementioned typical Tokamak magnetic field. 

\subsection{Banana Orbit} 

\begin{figure}[tbp]
\centering
\begin{subfigure}[tbp]{0.49\textwidth}
  \includegraphics[width=\textwidth]{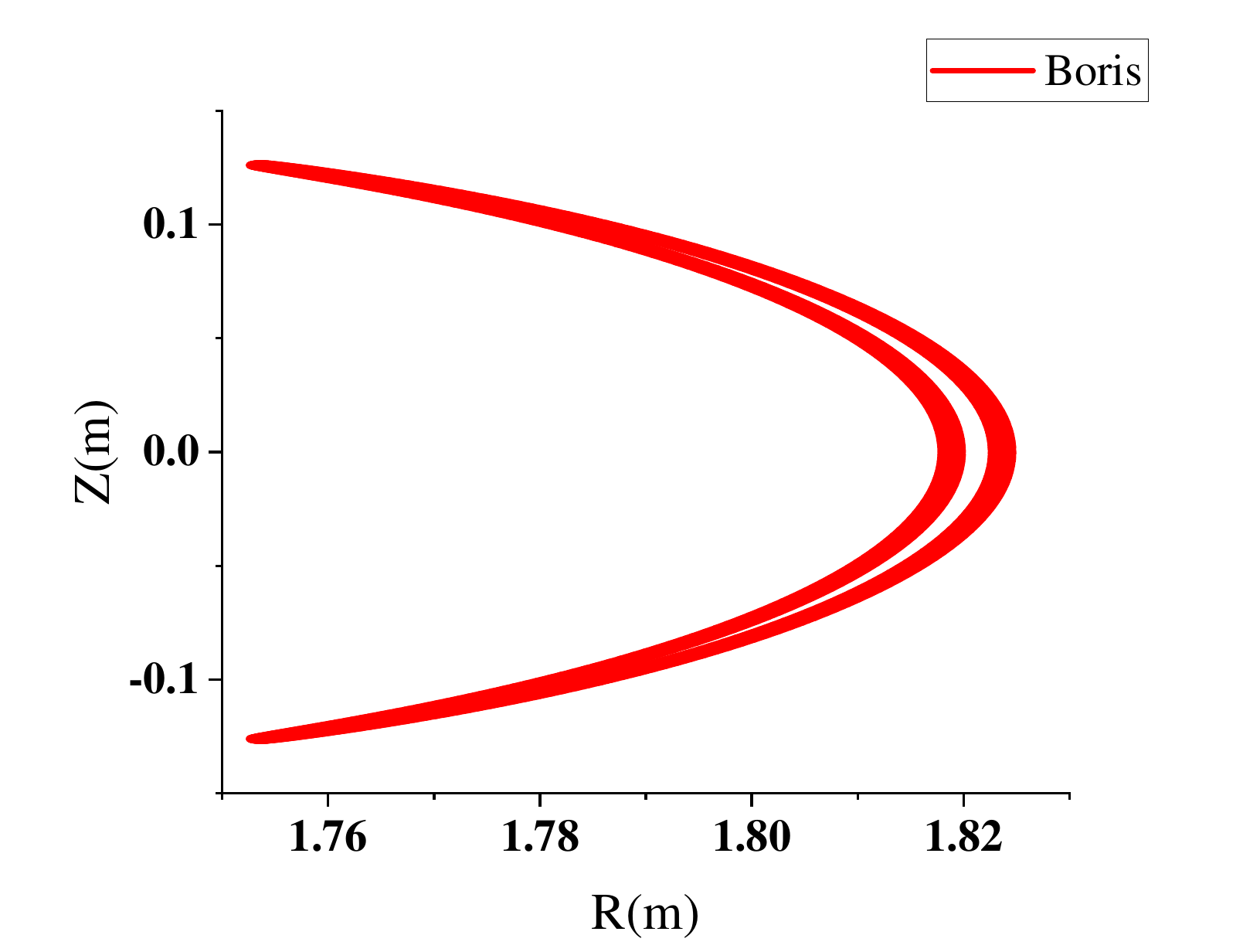}
  \caption{}
  \label{fig 1-a}
\end{subfigure}
\hfill 
\begin{subfigure}[tbp]{0.49\textwidth}
  \includegraphics[width=\textwidth]{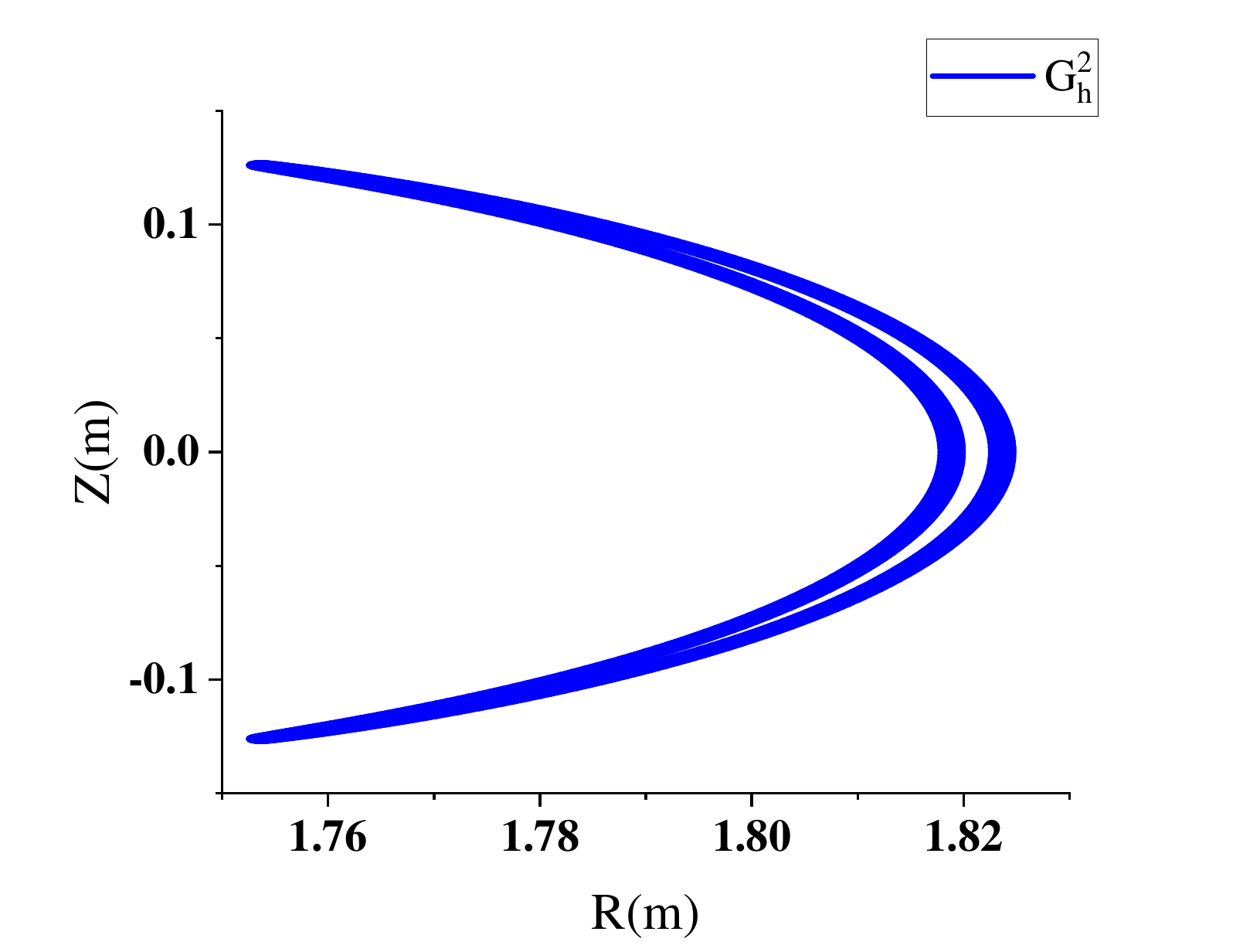}
  \caption{}
  \label{fig 1-b}
\end{subfigure}
\hfill 
\begin{subfigure}[tbp]{0.49\textwidth}
  \includegraphics[width=\textwidth]{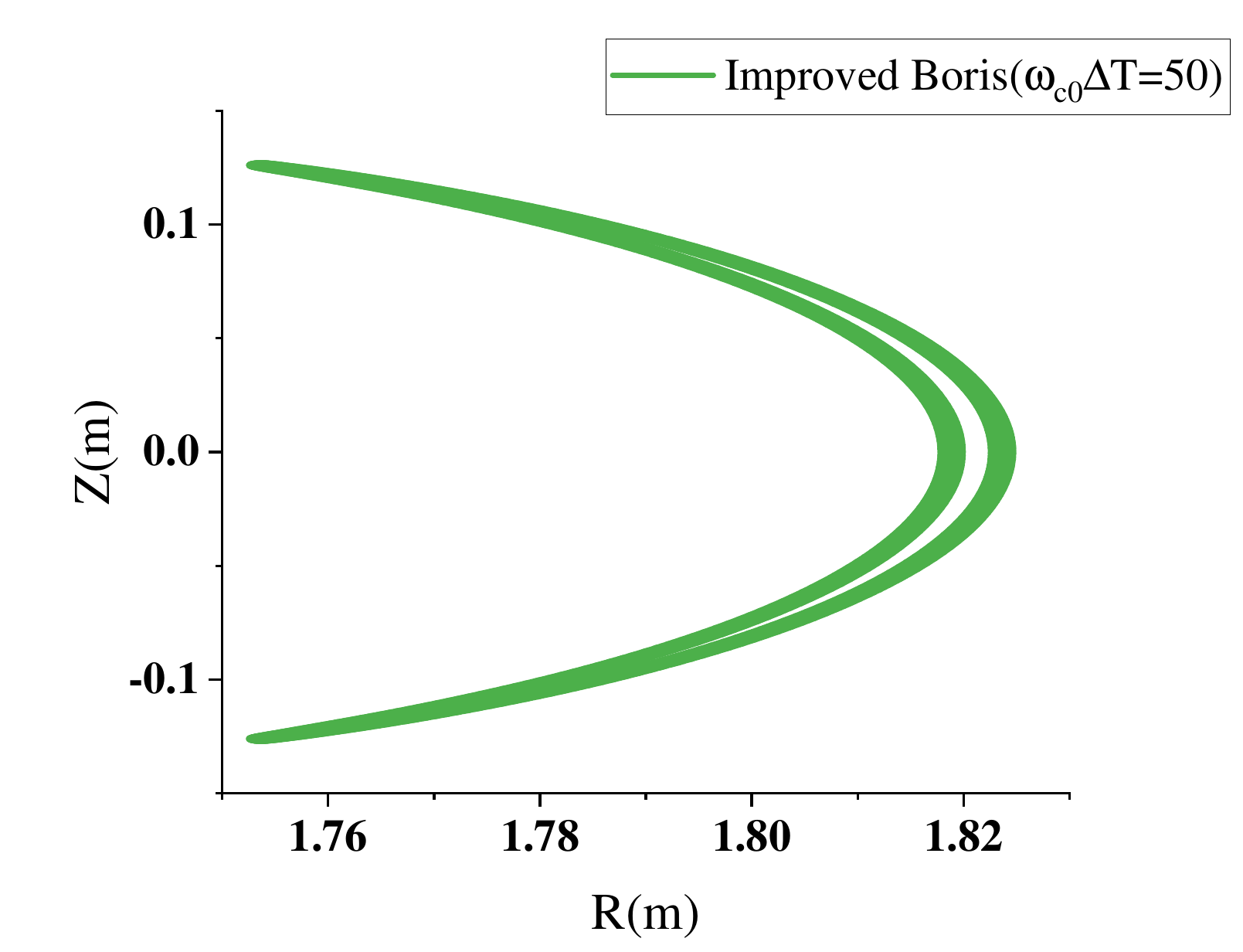}
  \caption{}
  \label{fig 1-c}
\end{subfigure}
\caption{Numerical results on the $(R,z)$ plane with initial conditions of banana orbit. The time step size is $\omega_{c0} \Delta t=0.1$, and the time integration interval is $[0,T_0]$ with $\omega_{c0}T_0=2.54 \times 10^4$ which is approximately one period of the slow-scale motions (banana period). The banana orbit is correctly obtained by all algorithms.}
\label{fig 1}
\end{figure}

~~~~The impact of the electric field is omitted in this case, i.e. $\vec E=(0,0,0)^T$. The initial velocity is {\color{blue}$\vec v_0=(0,2\times 10^4 \ m/s,2\times 10^5 \ m/s)^T$} and the initial position is {\color{blue}$\vec r_0=(R_0+0.25a,0,0)^T=(1.82\ m,0,0)^T$}. Under these conditions, the projection of the particle’s trajectory onto the $(R,z)$ plane {\color{blue}(where $R=\sqrt{x^2+y^2}$)} theoretically forms a closed banana orbit. All three algorithms are implemented with a relatively large time step size of $\omega_{c0} \Delta t=0.1$, and the time integration interval is $[0,T_0],\omega_{c0} T_0 =2.54\times 10^4$. The recalibration period of the improved Boris algorithm, $\Delta T$, is set to $\omega_{c0} \Delta T=50$, which remains constant throughout the section. The numerical results of the banana orbit are shown in Figure \ref{fig 1}. All algorithms correctly reproduce the trajectory of the trapped particle, and the results appear indistinguishable. 

\begin{figure}[tbp]
\centering
\begin{subfigure}[tbp]{0.49\textwidth}
  \includegraphics[width=\textwidth]{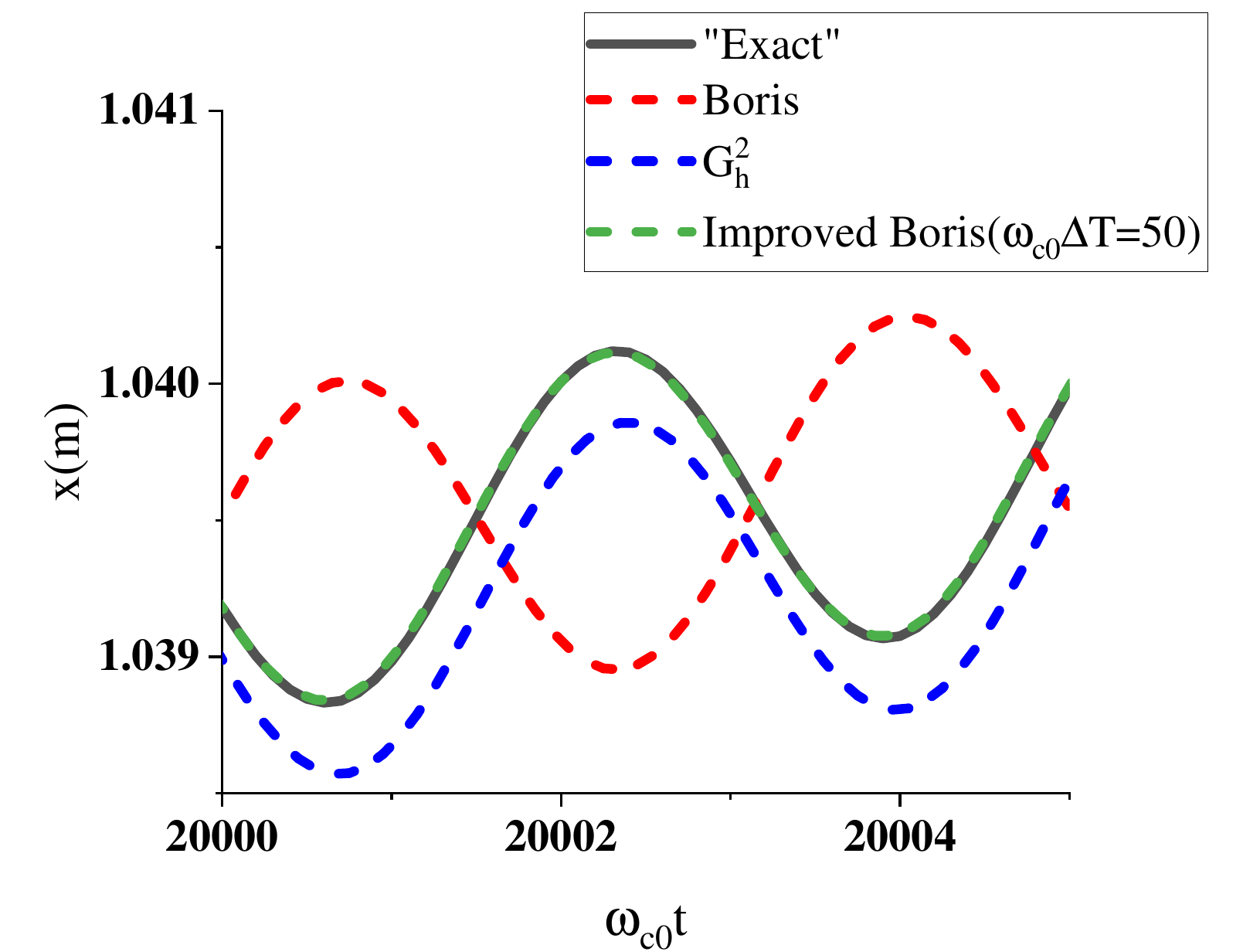}
  \caption{$x$ in [20000,20005]}
  \label{fig 2-a}
\end{subfigure}
\hfill 
\begin{subfigure}[tbp]{0.49\textwidth}
  \includegraphics[width=\textwidth]{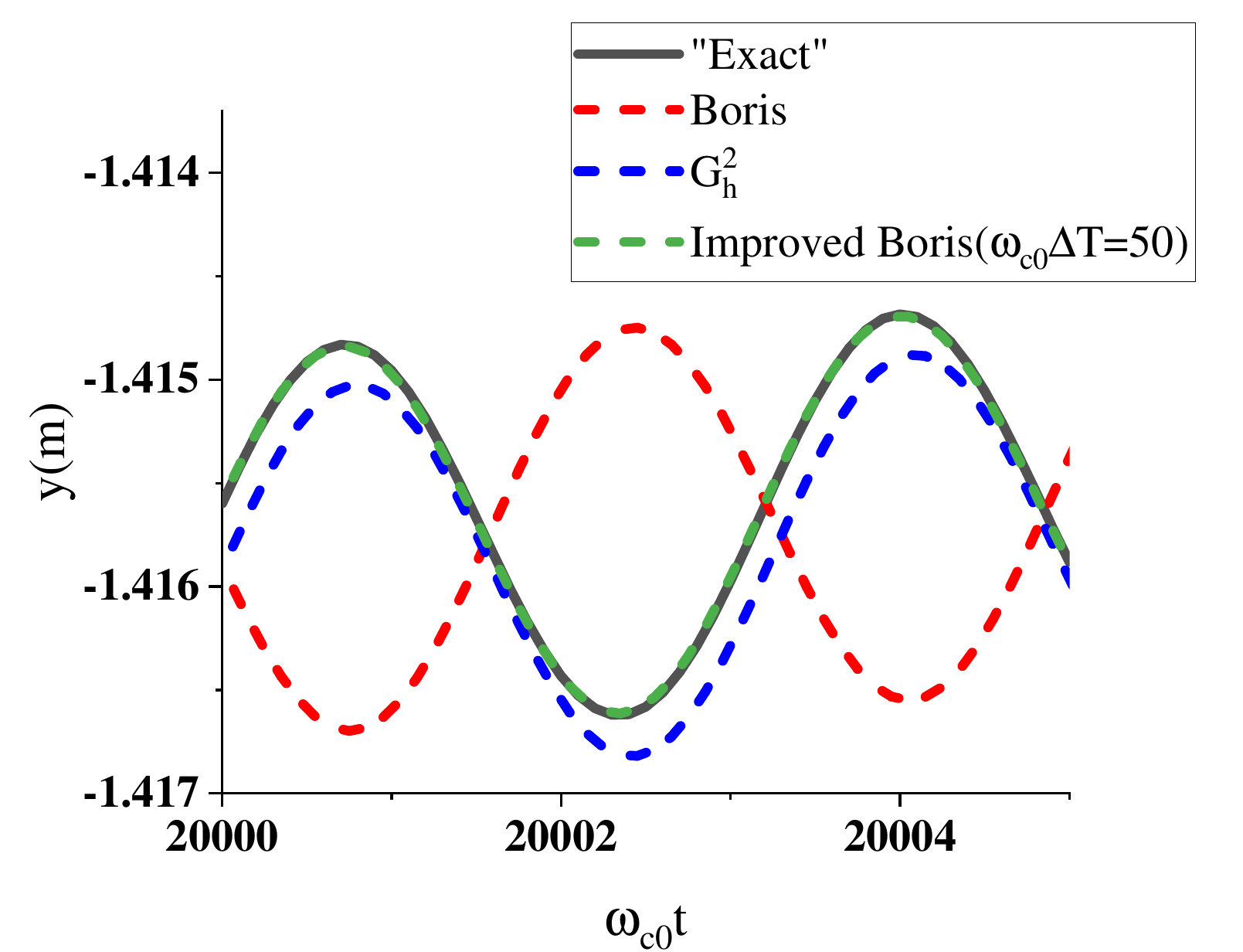}
  \caption{$y$ in [20000,20005]}
  \label{fig 2-b}
\end{subfigure}
\hfill 
\begin{subfigure}[tbp]{0.49\textwidth}
  \includegraphics[width=\textwidth]{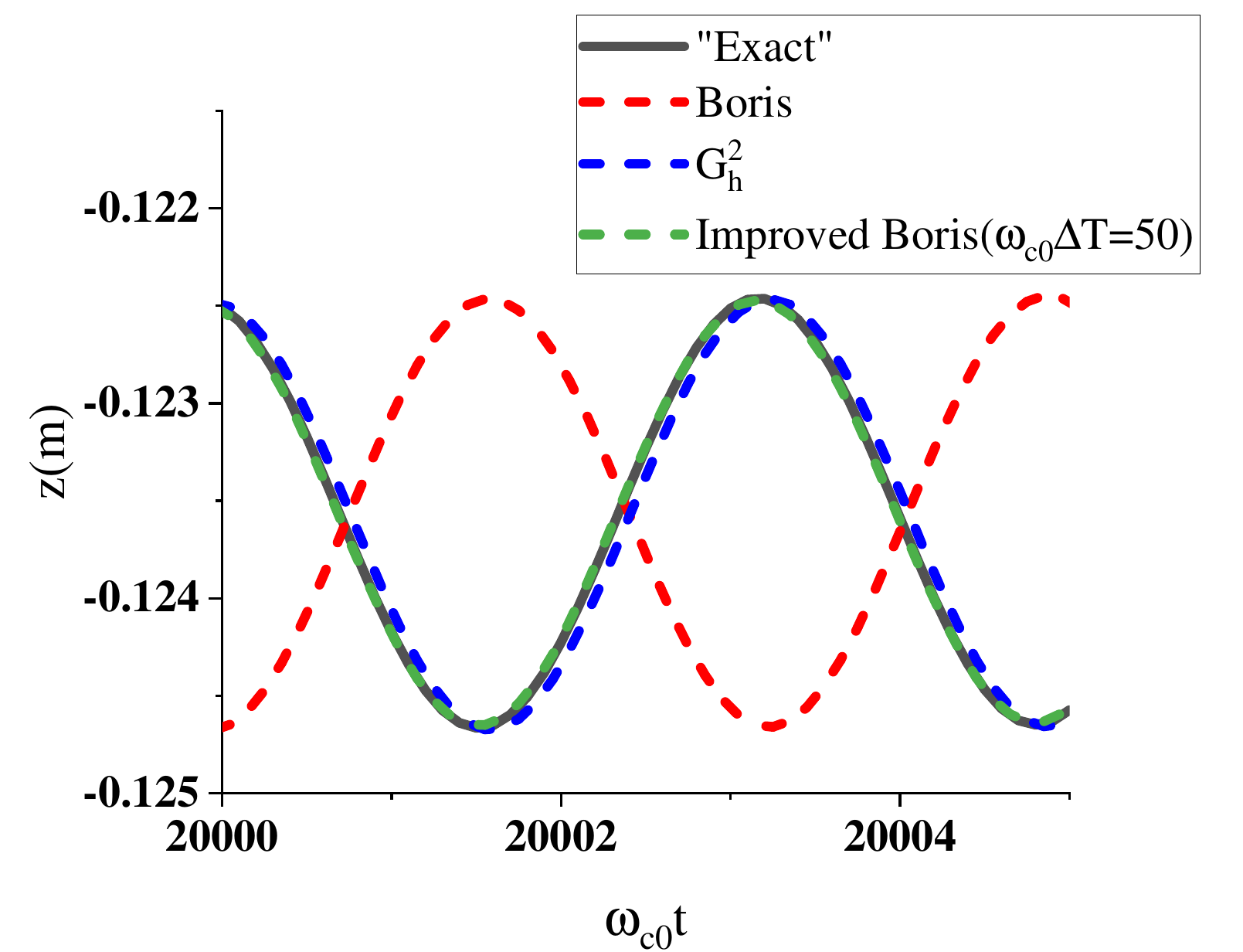}
  \caption{$z$ in [20000,20005]}
  \label{fig 2-c}
\end{subfigure}
\hfill 
\begin{subfigure}[tbp]{0.49\textwidth}
  \includegraphics[width=\textwidth]{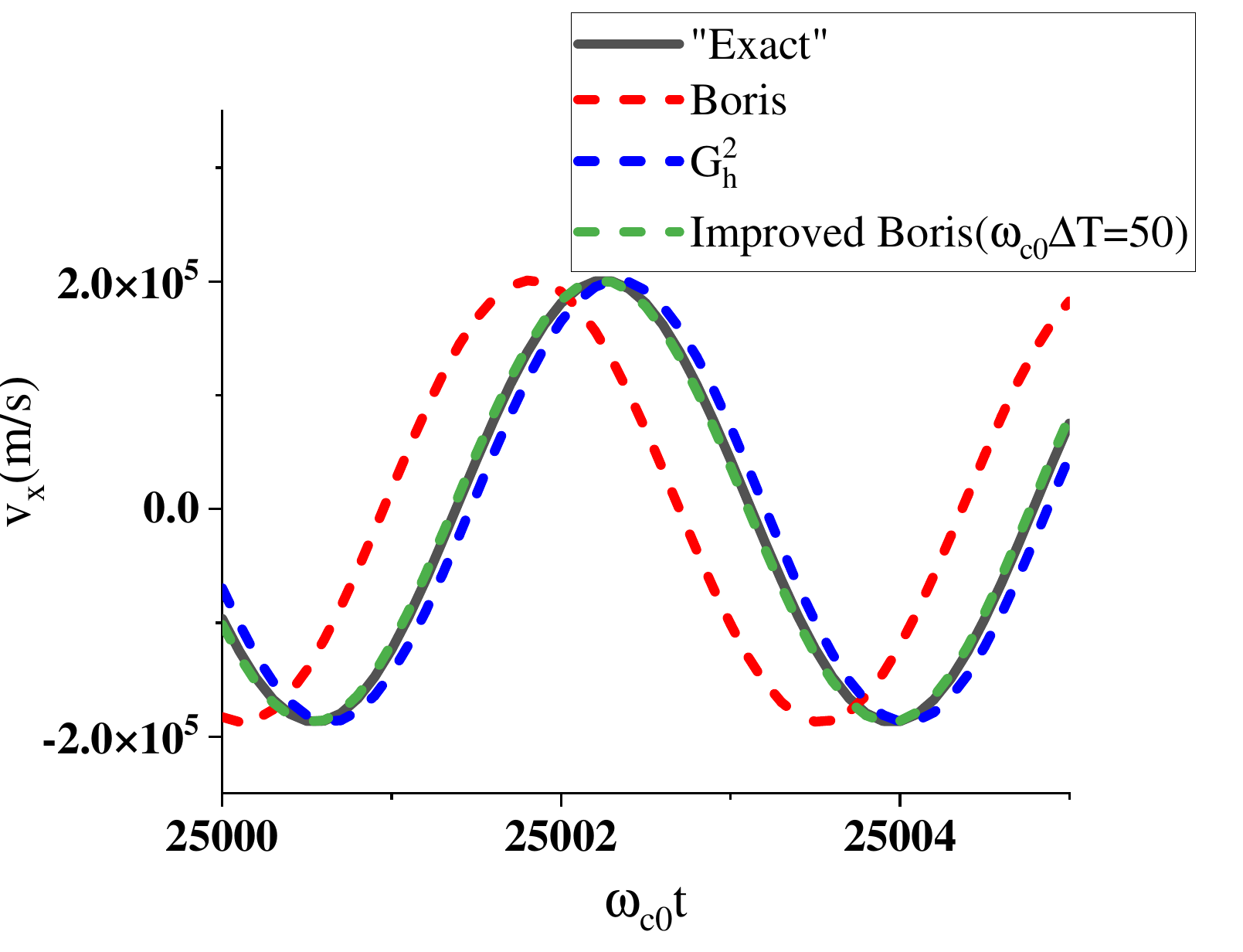}
  \caption{$v_x$ in [25000,25005]}
  \label{fig 2-d}
\end{subfigure}
\hfill 
\begin{subfigure}[tbp]{0.49\textwidth}
  \includegraphics[width=\textwidth]{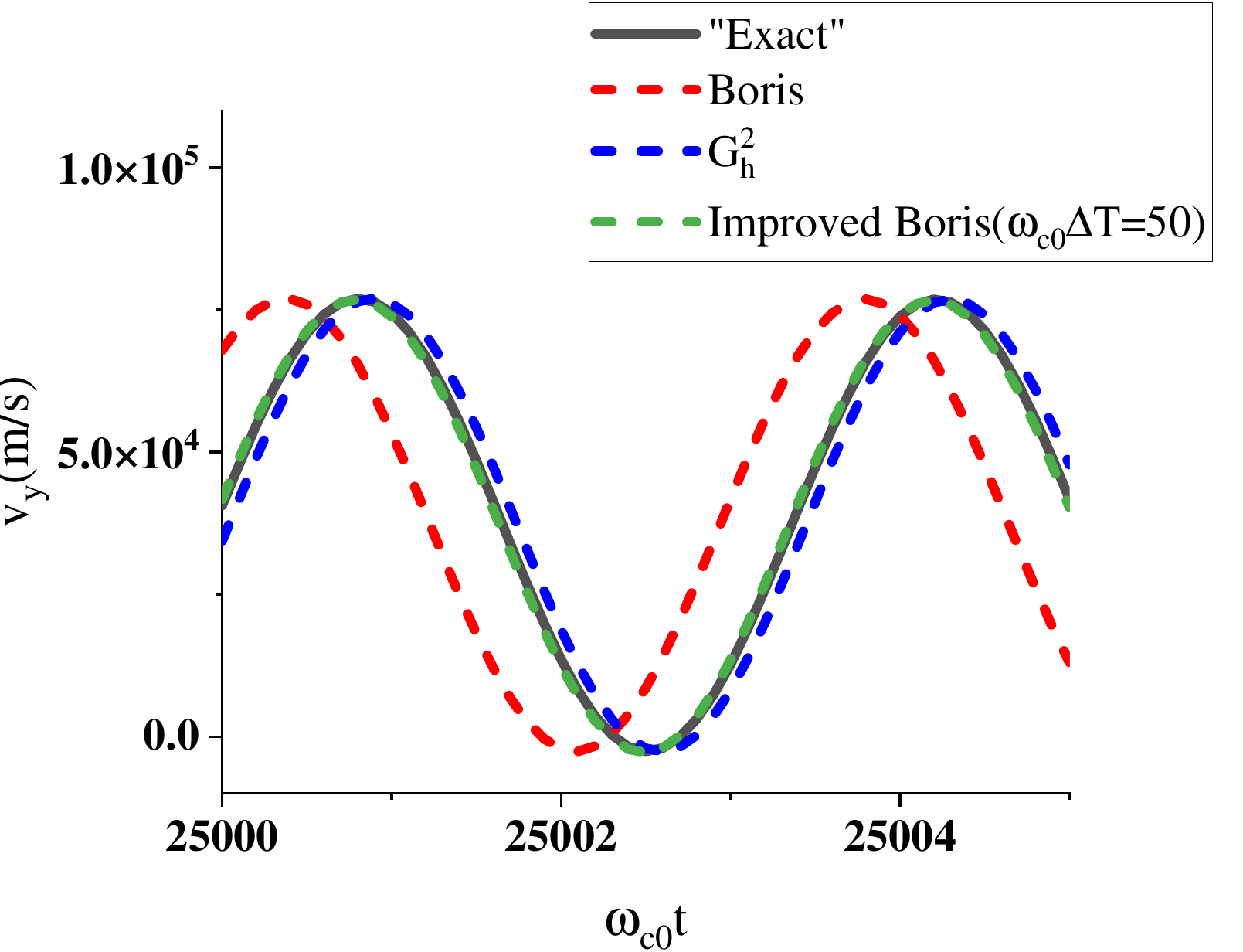}
  \caption{$v_y$ in [25000,25005]}
  \label{fig 2-e}
\end{subfigure}
\hfill 
\begin{subfigure}[tbp]{0.49\textwidth}
  \includegraphics[width=\textwidth]{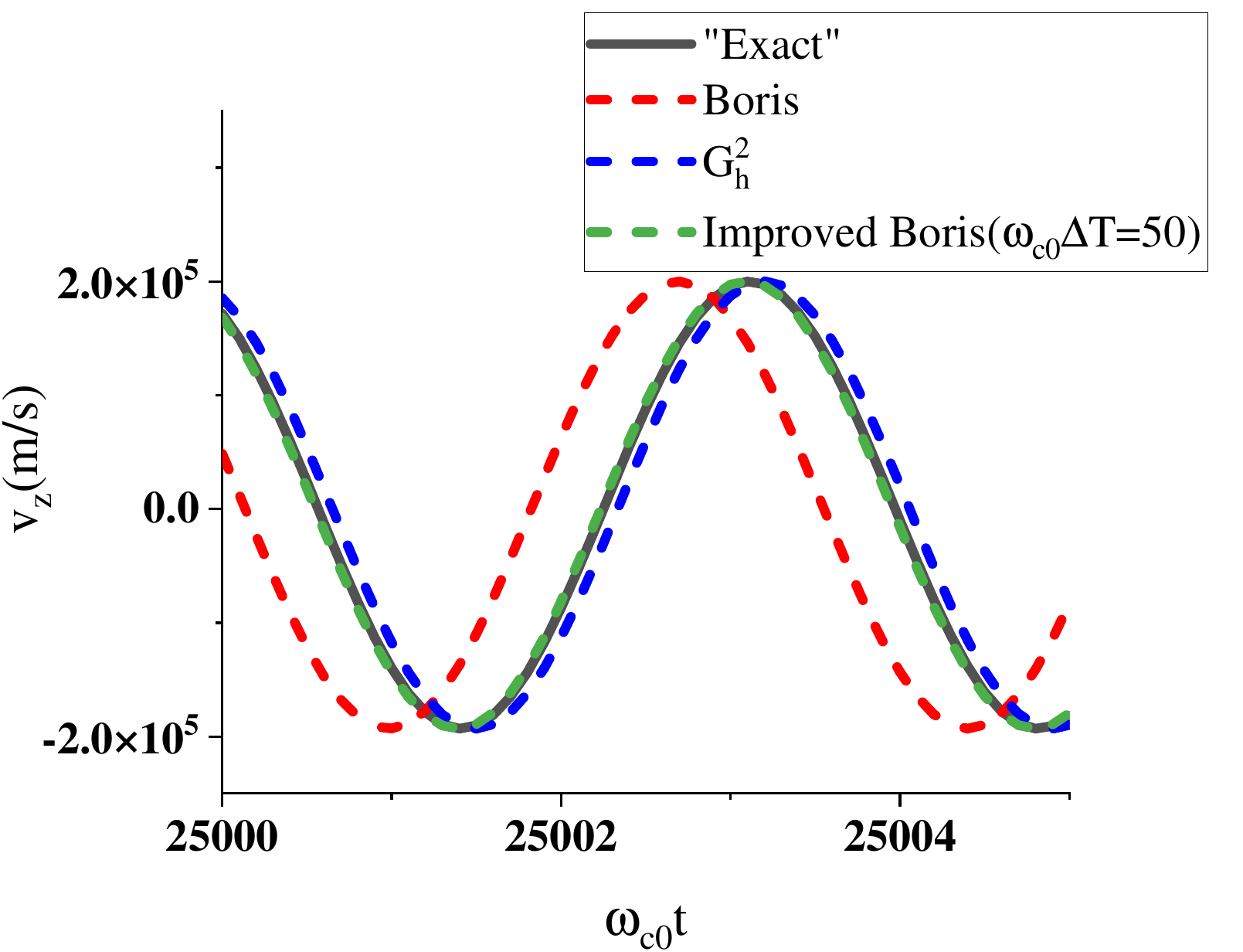}
  \caption{$v_z$ in [25000,25005]}
  \label{fig 2-f}
\end{subfigure}
\caption{Time-dependent numerical results of the banana orbit in selected time intervals with a time step size of $\omega_{c0} \Delta t=0.1$. (a)-(c): $\vec r$ in [20000,20005]. (d)-(f): $\vec v$ in [25000,25005]. The Boris algorithm {\color{blue}(red dashed lines)} and $G_h^2$ {\color{blue}(blue dashed lines)} introduce significant errors, while the improved Boris algorithm {\color{blue}(green dashed lines)} demonstrates much higher accuracy. As observed in the figure, its numerical solution is almost indistinguishable from the "exact" solution {\color{blue}(black solid lines)}.}
\label{fig 2}
\end{figure}

To further distinguish the performance of the three algorithms, we examine the time-dependent numerical results of the position $\vec{r}$ and velocity $\vec{v}$ over selected intervals. Shown in Figure \ref{fig 2} is the time-dependent numerical results of the banana orbit by all algorithms with $\omega_{c0} \Delta t=0.1$ in selected time intervals of an equal length of 5, compared with the “exact” solutions obtained by an extremely small time step size {\color{blue}($\omega_{c0} \Delta t=10^{-5}$, which is practically unattainable in actual numerical simulations, and all algorithms generate identical results in this case. Here we select the result derived by the Boris algorithm)}. The numerical solutions of the position variables $\vec r$ in the time interval $[20000,20005]$ are displayed in sub-figures (a), (b) and (c). It is evident that, in comparison to the "exact" solution, the numerical result obtained from the Boris algorithm  exhibits a distinct phase discrepancy. In contrast, while the $G_h^2$ solutions aligns more closely in terms of phase, a noticeable overall shift in the particle trajectory is observed. The improved Boris algorithm yields superior results, with the numerical solution closely overlapping with the "exact" solution, as observed in the figures. This favorable outcome is consistently maintained in the numerical solutions of the velocity variables $\vec v$ in the time interval [25000,25005], as shown in sub-figures (d), (e) and (f). The Boris algorithm still maintains a significant phase error. As for $G_h^2$, due to the overall trajectory shift mentioned earlier, a slight inaccuracy in the magnetic field is introduced, resulting in a minor phase discrepancy.       

{\color{blue}In addition to the zoomed-in view of the short-term behavior presented in Figure \ref{fig 2}, the long-term behavior of the numerical solutions is also examined by evaluating the evolution of relative errors throughout the entire calculation process, as shown in Figure \ref{fig 2-1}. Here, the relative errors in the $m-th$ time step $e_{\vec r}^m$ and $e_{\vec v}^m$ are defined by 
$$
e_{\vec r}^m = \sqrt{\frac{|\vec r_m^{"exact"}-\vec r_m^{numerical}|^2}{|\vec r_m^{"exact"}|^2}}
\eqno{(16.a)}
$$
$$
e_{\vec v}^m = \sqrt{\frac{|\vec v_m^{"exact"}-\vec v_m^{numerical}|^2}{|\vec v_m^{"exact"}|^2}}
\eqno{(16.b)}
$$
In Figure \ref{fig 2-1}, the relative errors of the Boris algorithm display periodic fluctuations, with the amplitude of these oscillations remaining largely constant, consistent with its characteristic of exhibiting no trajectory drift but an inaccurate cyclotron phase. In contrast, $G_h^2$ displays initially small relative errors. However, as simulation time increases, the cumulative spatial drift in the particle trajectory leads to a steady rise in the error, resulting in a long-term error accumulation. And the improved Boris algorithm clearly outperforms both baseline methods, maintaining substantially lower relative errors over the entire simulation. 

\begin{figure}[tbp]
\centering
\begin{subfigure}[tbp]{0.49\textwidth}
  \includegraphics[width=\textwidth]{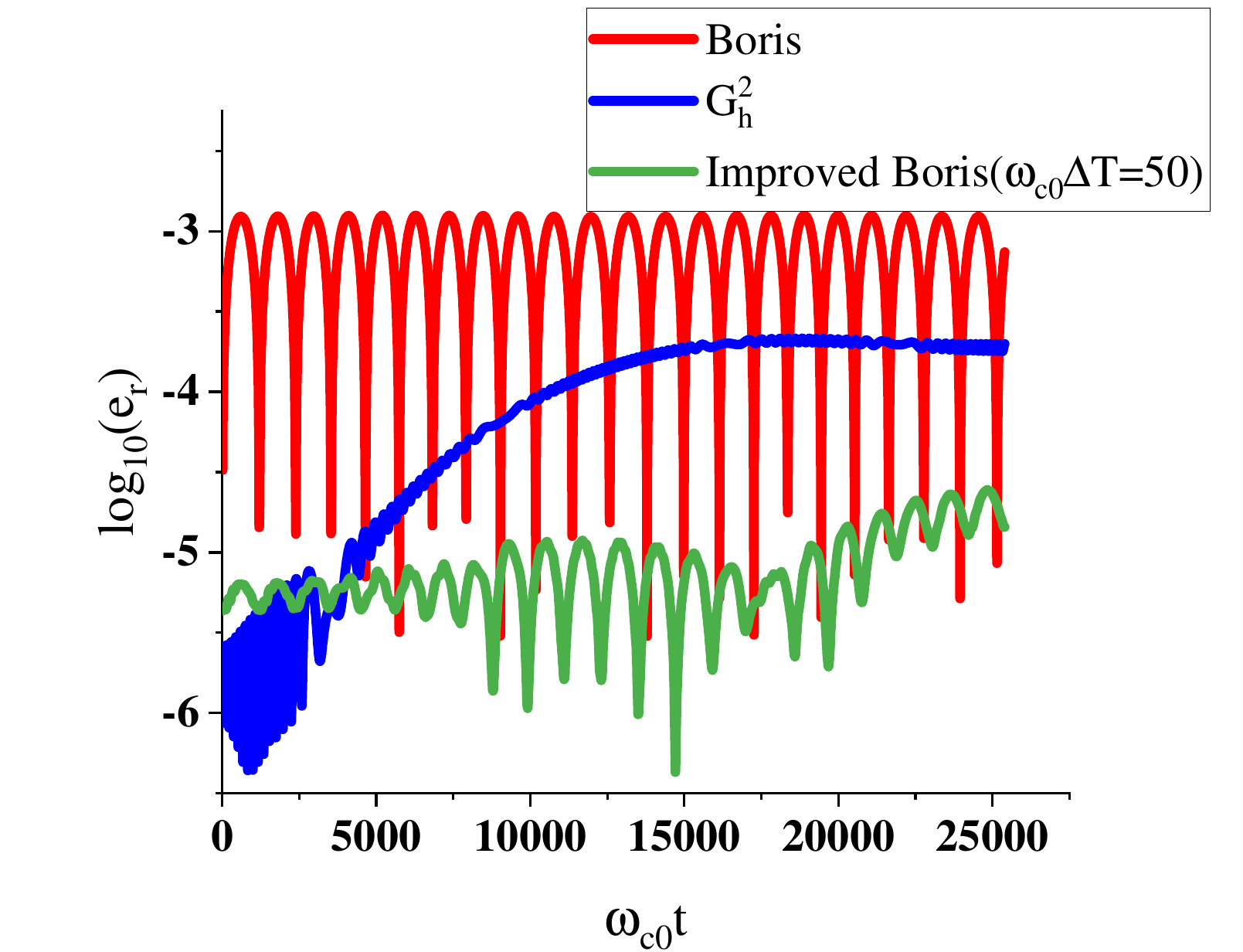}
  \caption{$e_{\vec r}$ in $[0,T_0]$}
  \label{fig 2-1-a}
\end{subfigure}
\hfill 
\begin{subfigure}[tbp]{0.49\textwidth}
  \includegraphics[width=\textwidth]{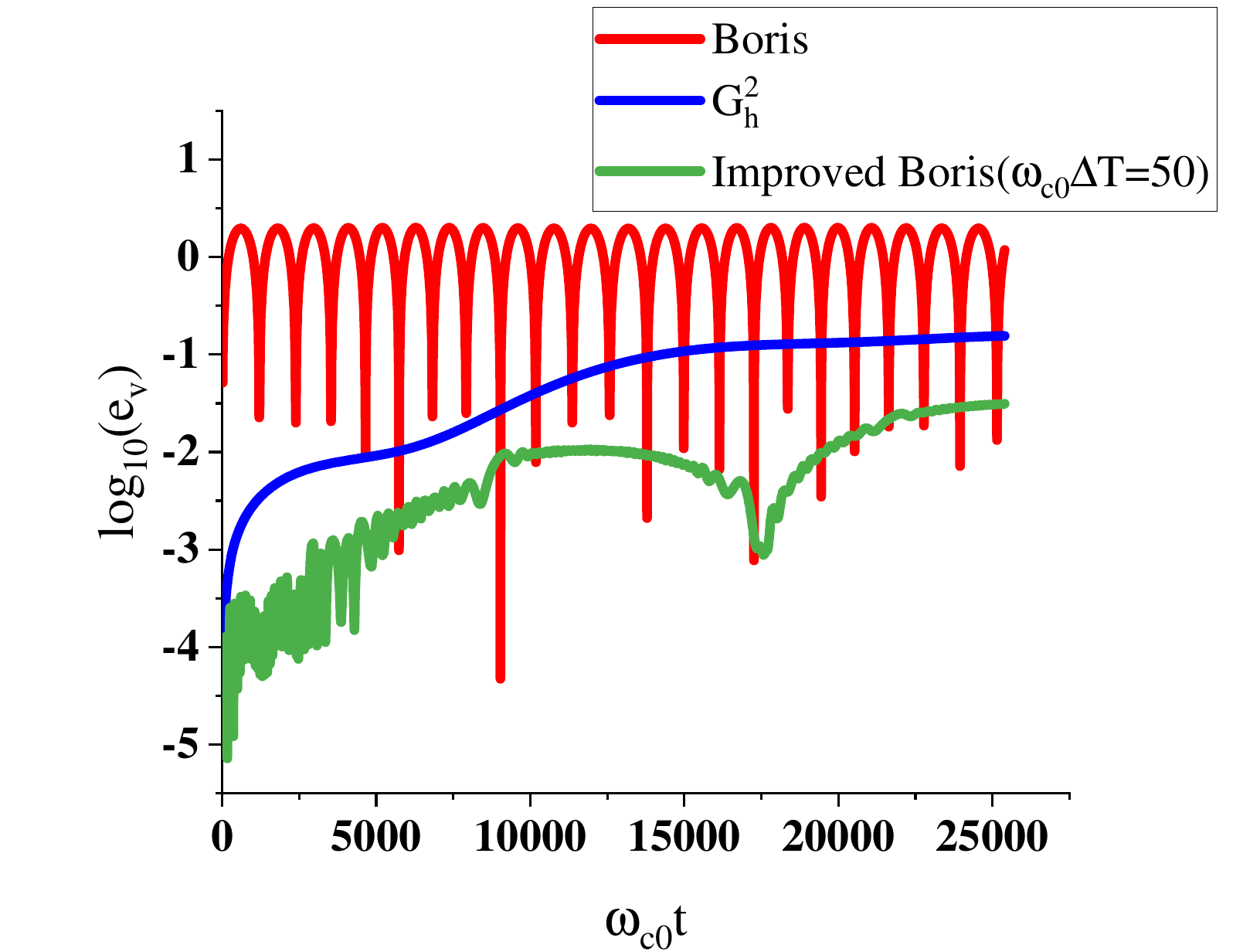}
  \caption{$e_{\vec v}$ in $[0,T_0]$}
  \label{fig 2-1-b}
\end{subfigure}
\caption{Time-dependent relative errors of $\vec r$ and $\vec v$ in $[0,T_0]$ by all algorithms.}
\label{fig 2-1}
\end{figure}
}

{\color{blue} In the absence of the electric field, the kinetic energy $E=\frac{1}{2} v^2$ is a constant of motion. Additionally, due to the slow spatial variation of the magnetic field, the magnetic moment $\mu=\frac{v_{\perp}^2}{2B}$ is an adiabatic invariant. The numerical conservation of these physical quantities is illustrated in Figure \ref{fig 3-1}, by similarly treating the relative errors as a function of integration time. In this case, since all three algorithms are theoretically capable of strictly conserving the magnitude of velocity $v$, the relative errors of kinetic energy $E$ remain at the level of machine precision, as shown in Figure \ref{fig 3-1-a}. Meanwhile, Figure \ref{fig 3-1-b} demonstrates that all three methods are capable of effectively preserving the magnetic moment $\mu$ over time. 

\begin{figure}[tbp]
\centering
\begin{subfigure}[tbp]{0.49\textwidth}
  \includegraphics[width=\textwidth]{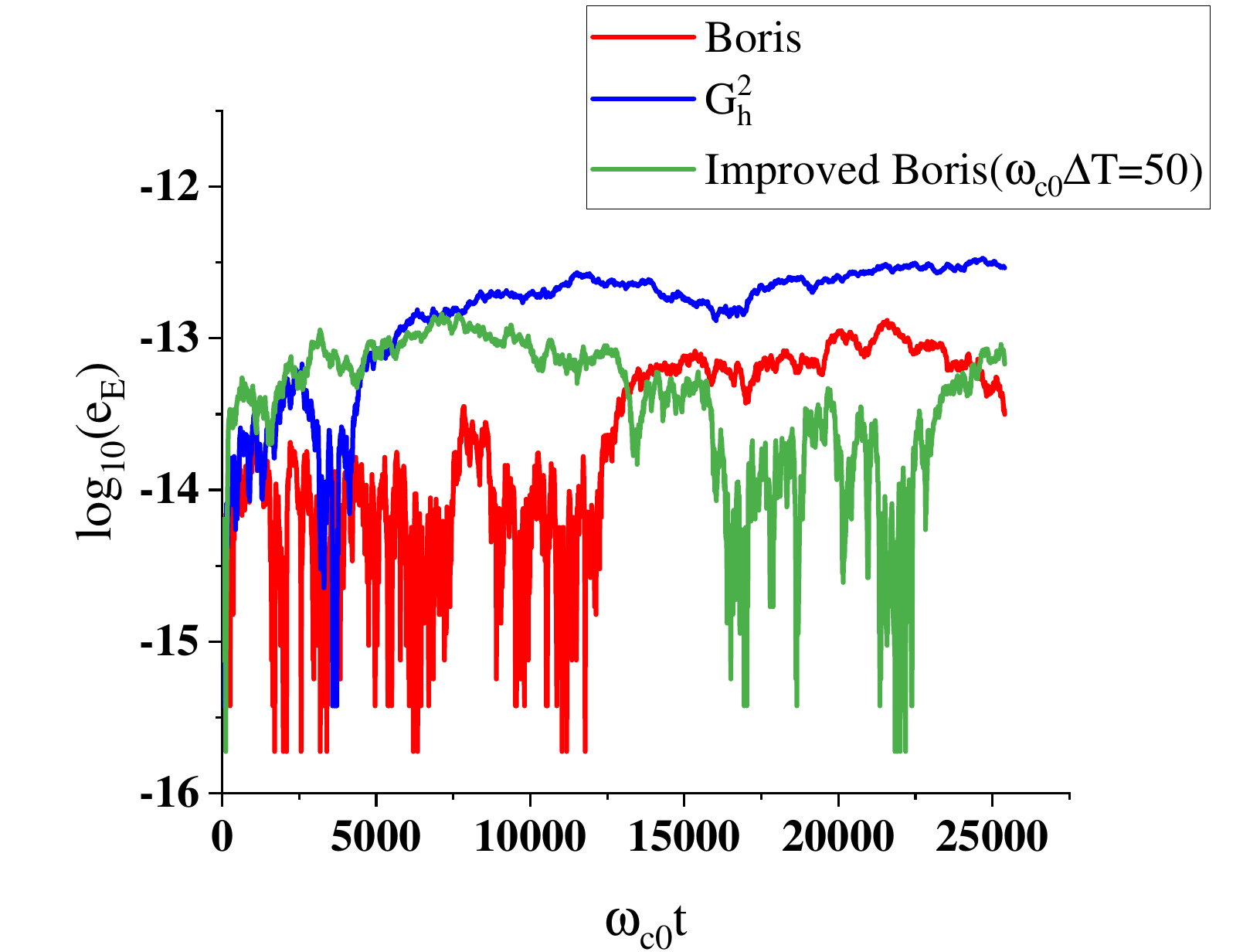}
  \caption{relative errors of energy in $[0,T_0]$}
  \label{fig 3-1-a}
\end{subfigure}
\hfill 
\begin{subfigure}[tbp]{0.49\textwidth}
  \includegraphics[width=\textwidth]{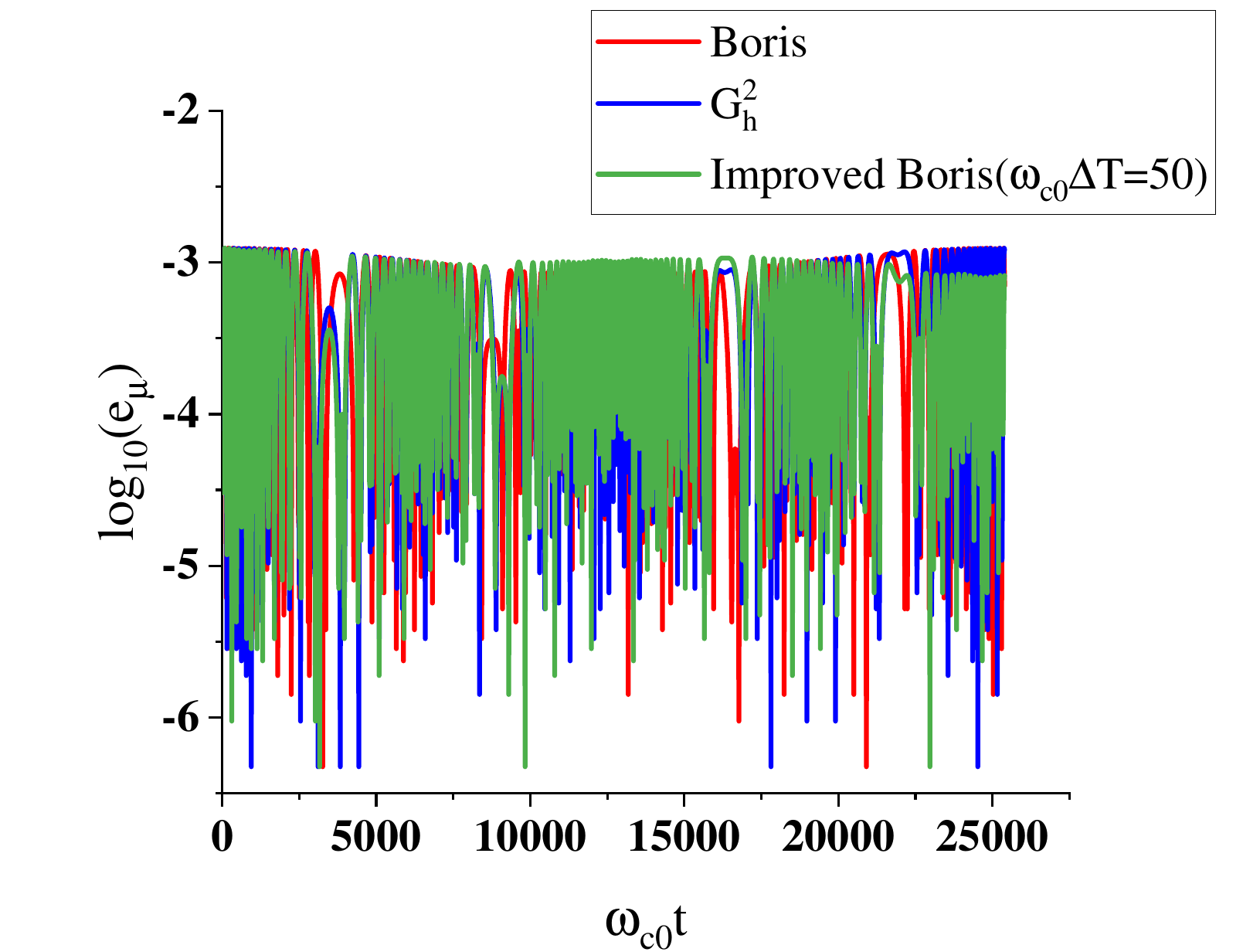}
  \caption{relative errors of magnetic moment in $[0,T_0]$}
  \label{fig 3-1-b}
\end{subfigure}
\caption{Time-dependent relative errors of energy and magnetic moment in $[0,T_0]$ by all algorithms.}
\label{fig 3-1}
\end{figure}
}

\begin{figure}[tbp]
\centering
\begin{subfigure}[tbp]{0.49\textwidth}
  \includegraphics[width=\textwidth]{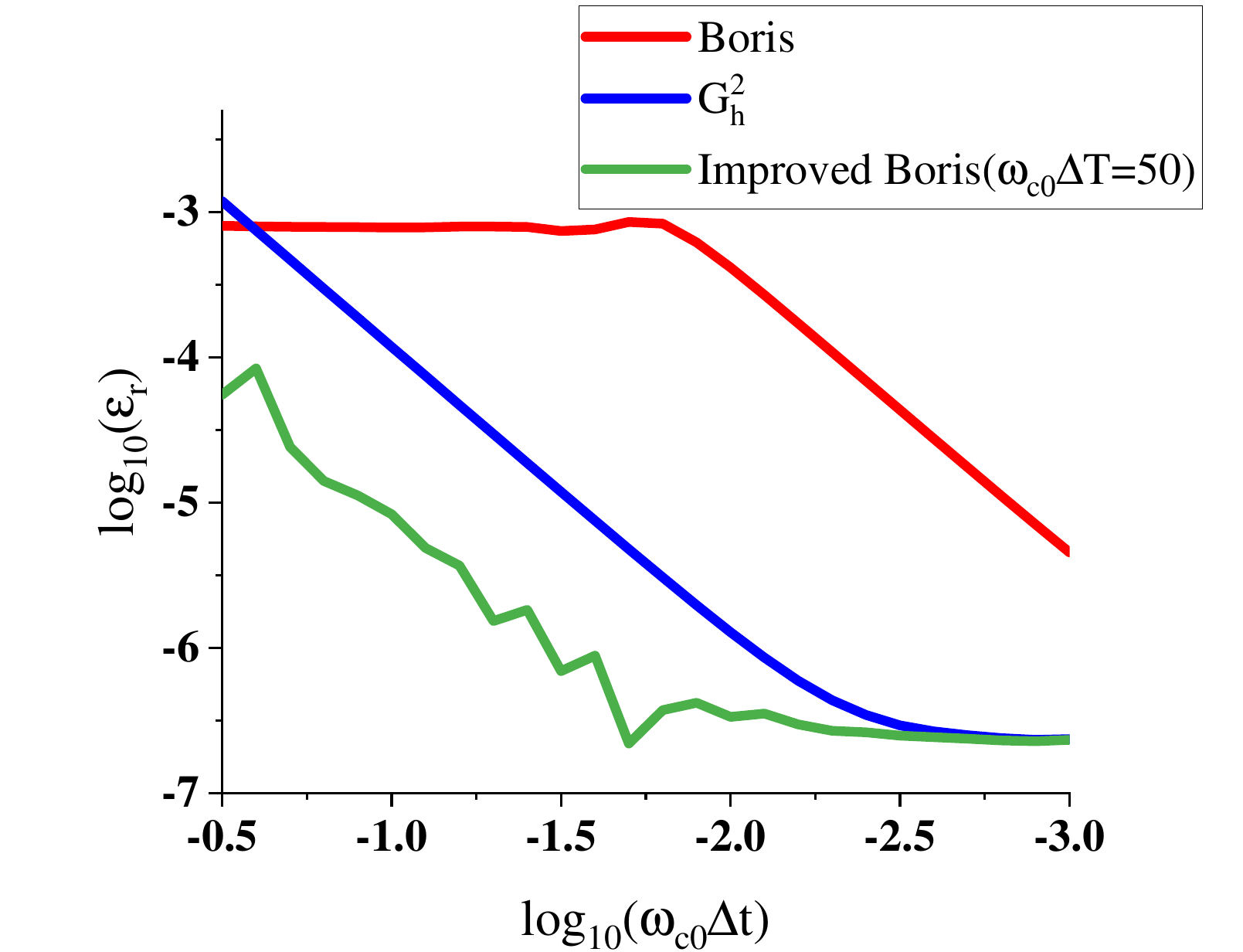}
  \caption{average relative errors of position}
  \label{fig 3-a}
\end{subfigure}
\hfill 
\begin{subfigure}[tbp]{0.49\textwidth}
  \includegraphics[width=\textwidth]{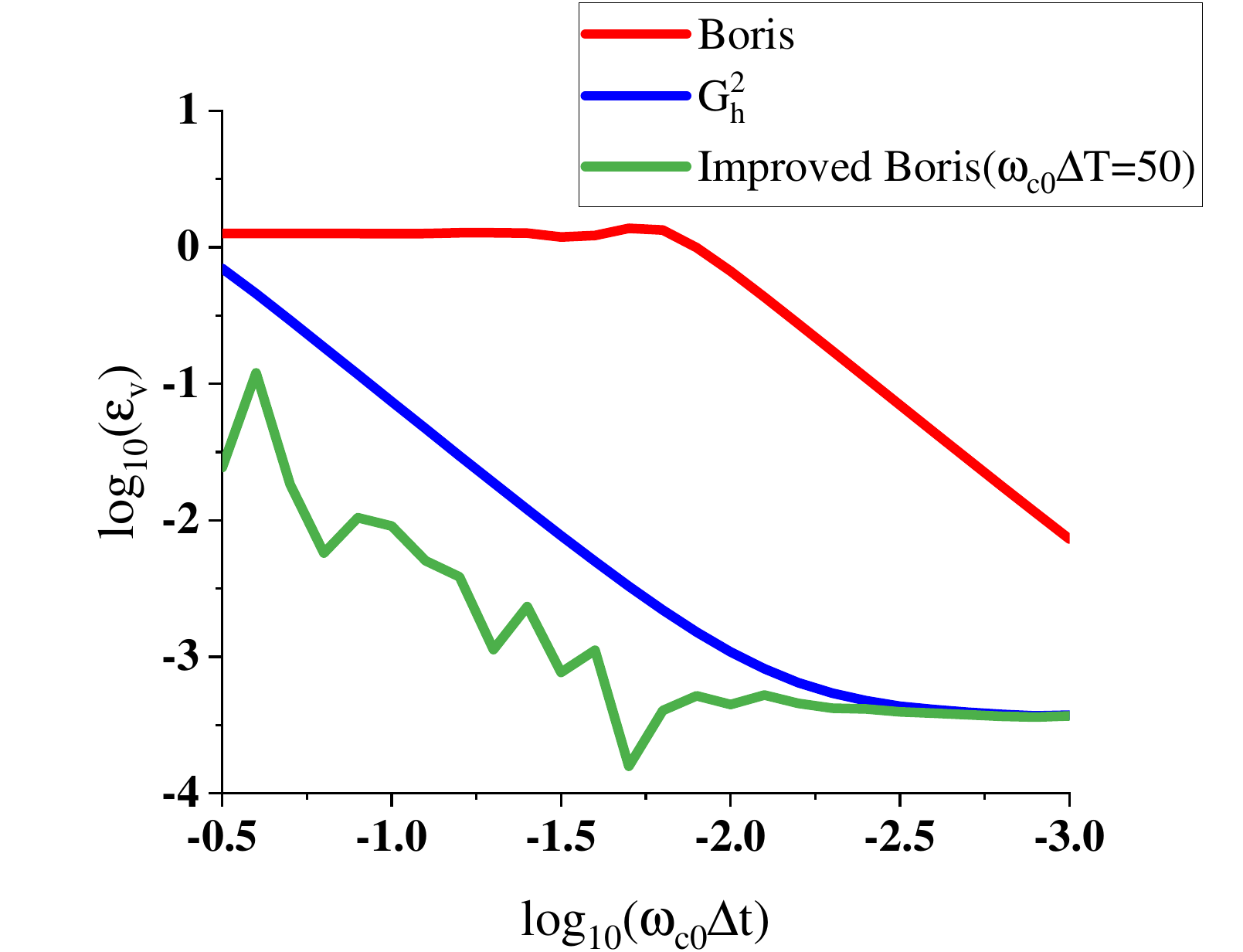}
  \caption{average relative errors of velocity}
  \label{fig 3-b}
\end{subfigure}
\caption{Average relative errors of $\vec r$ and $\vec v$ as functions of time step size $\Delta t$ by all algorithms. The accuracy of the improved Boris algorithm significantly exceeds that of the other two volume-preserving algorithms.}
\label{fig 3}
\end{figure}

To provide a more intuitive comparison of the accuracy of these algorithms, Figure \ref{fig 3} presents the average relative errors over the entire time integration interval $[0,T_0]$ as a function of time step size $\Delta t$. Here, the average relative errors are defined by
$$
\epsilon_{\vec r} = \frac{1}{N} \sum_{m=0}^{N-1} e_{\vec r}^m
\eqno{(17.a)}
$$
$$
\epsilon_{\vec v} = \frac{1}{N} \sum_{m=0}^{N-1} e_{\vec v}^m
\eqno{(17.b)}
$$
with $N=\frac{T_0}{\Delta t}$ the total number of time grids. It can be observed that the convergence curves of both VPAs are remarkably smooth, with the slope of the linear region approaching 2, indicating second-order convergence rate for both algorithms, while $G_h^2$ converges noticeably faster than the Boris algorithm. Although the convergence curve of the improved Boris algorithm appears less smooth, it achieves an improvement of approximately one order of magnitude over $G_h^2$, which is the more accurate scheme of the two baseline algorithms.   

\begin{figure}[tbp]
\centering
\begin{subfigure}[tbp]{0.49\textwidth}
  \includegraphics[width=\textwidth]{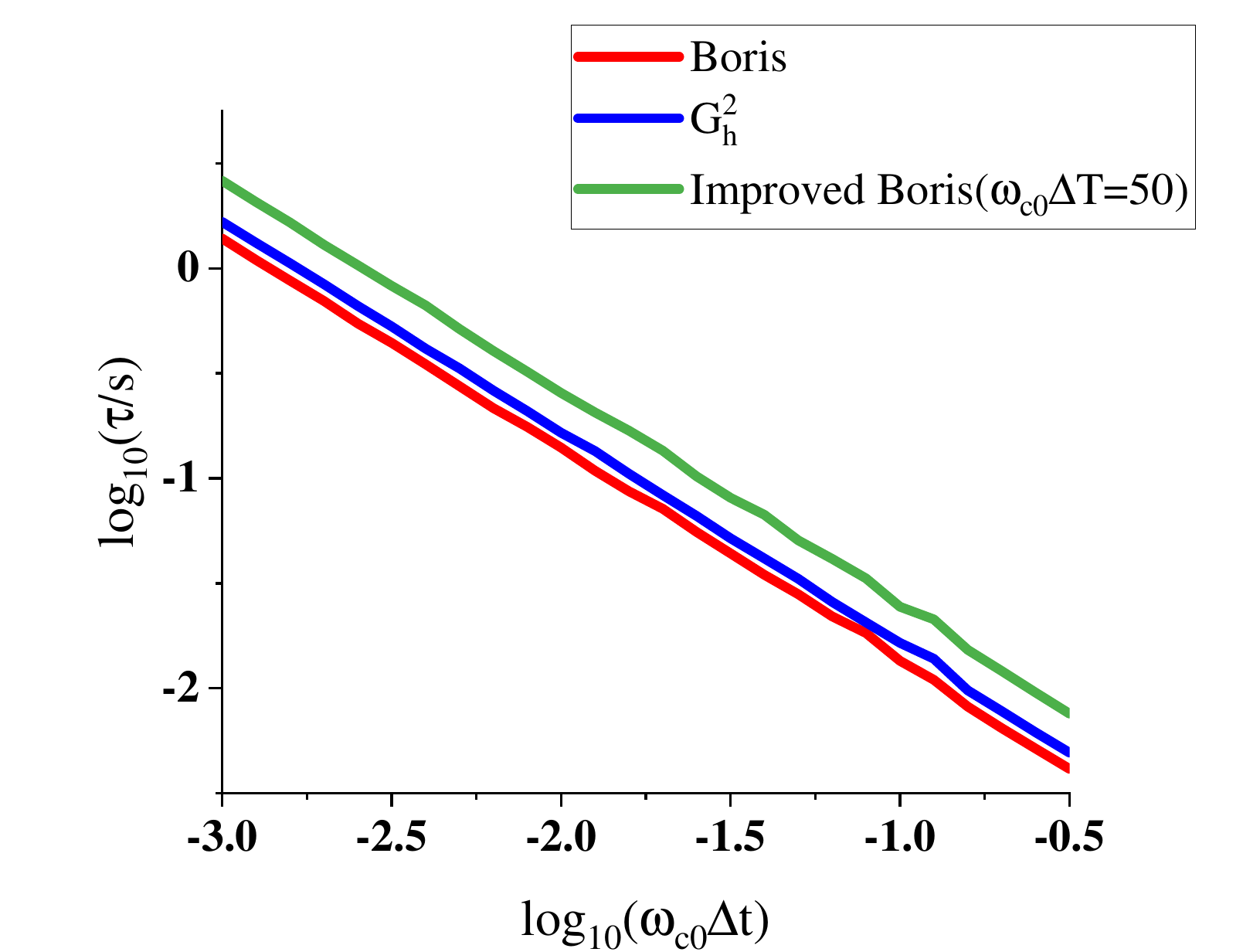}
  \caption{computational time}
  \label{fig 4-a}
\end{subfigure}
\hfill 
\begin{subfigure}[tbp]{0.49\textwidth}
  \includegraphics[width=\textwidth]{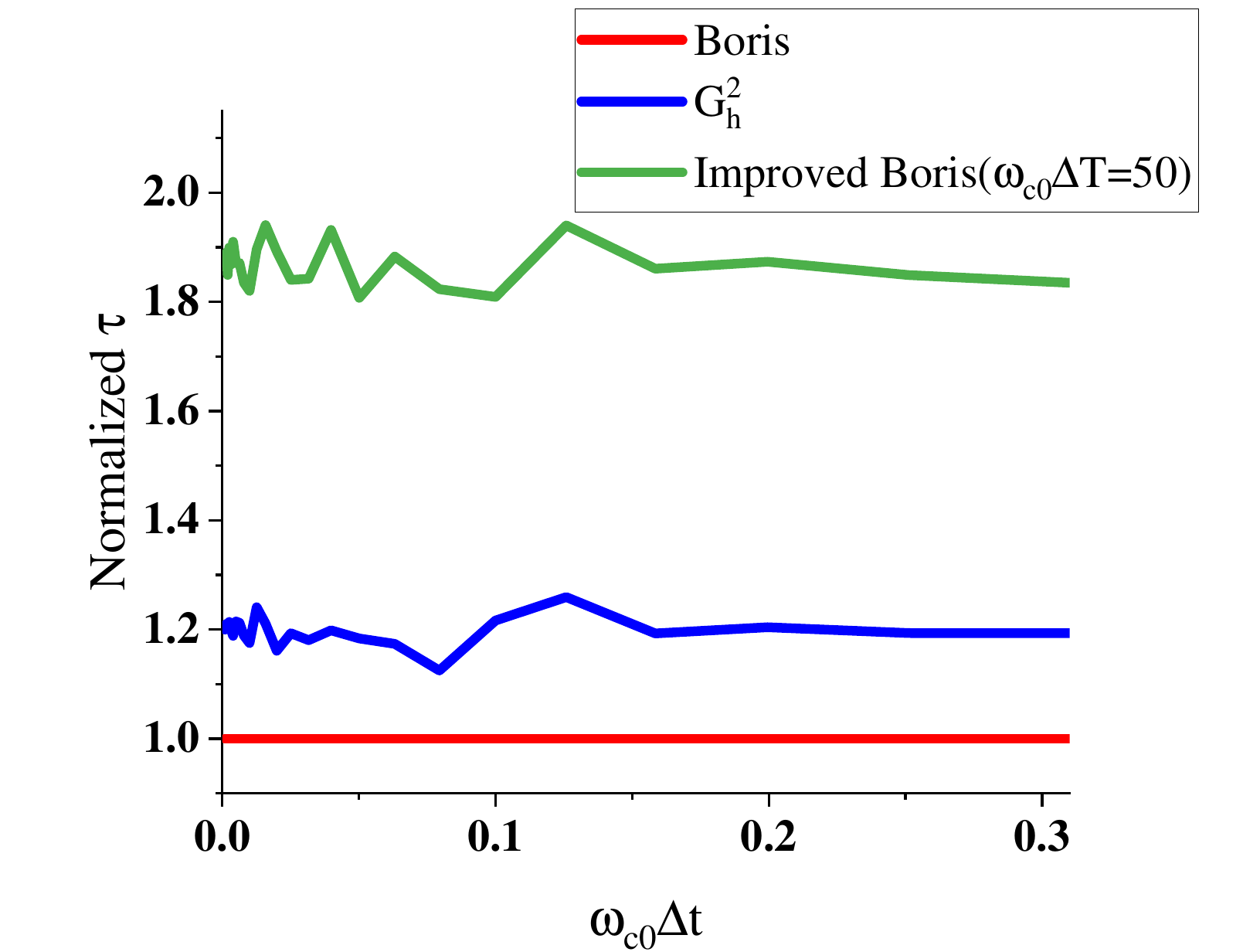}
  \caption{normalized computational time}
  \label{fig 4-b}
\end{subfigure}
\caption{Computational time of all algorithms.}
\label{fig 4}
\end{figure}

Compared in Figure \ref{fig 4} is the computational time $\tau$ {\color{blue}(measured in seconds and averaged over 100 repeated simulations)} of the three algorithms. It should be noted that the improved Boris algorithm is non-parallelized. Sub-figure (a) shows the absolute computation time under identical conditions. As expected, the Boris algorithm achieves the shortest runtime, followed by $G_h^2$, while the Improved Boris Algorithm exhibits the longest computation time due to its combined structure. In sub-figure (b), the computation times of all algorithms are normalized with respect to the Boris algorithm's computation time. This normalization reveals that the computation time for $G_h^2$ is approximately 20\% greater than that of the Boris algorithm, attributable to the need for trigonometric function calculations when computing the corresponding rotation matrix $R_k$. The non-parallelized improved Boris algorithm incurs a computational cost approximately 1.8 to 2.0 times that of the Boris algorithm, which is slightly lower than the sum of the individual runtimes of the Boris algorithm and $G_h^2$.  

\begin{figure}[tbp]
\centering
\begin{subfigure}[tbp]{0.49\textwidth}
  \includegraphics[width=\textwidth]{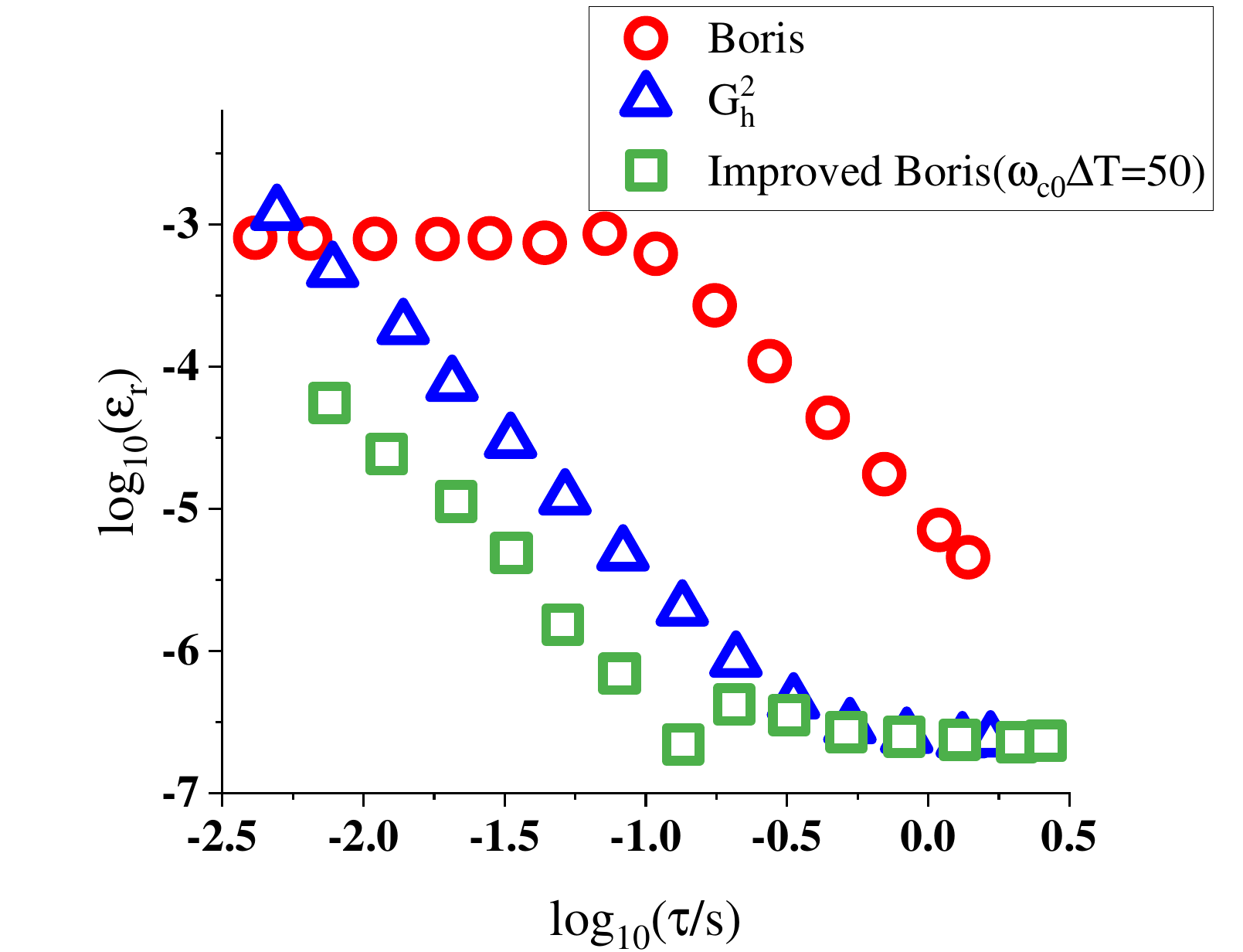}
  \caption{efficiency of position}
  \label{fig 5-a}
\end{subfigure}
\hfill 
\begin{subfigure}[tbp]{0.49\textwidth}
  \includegraphics[width=\textwidth]{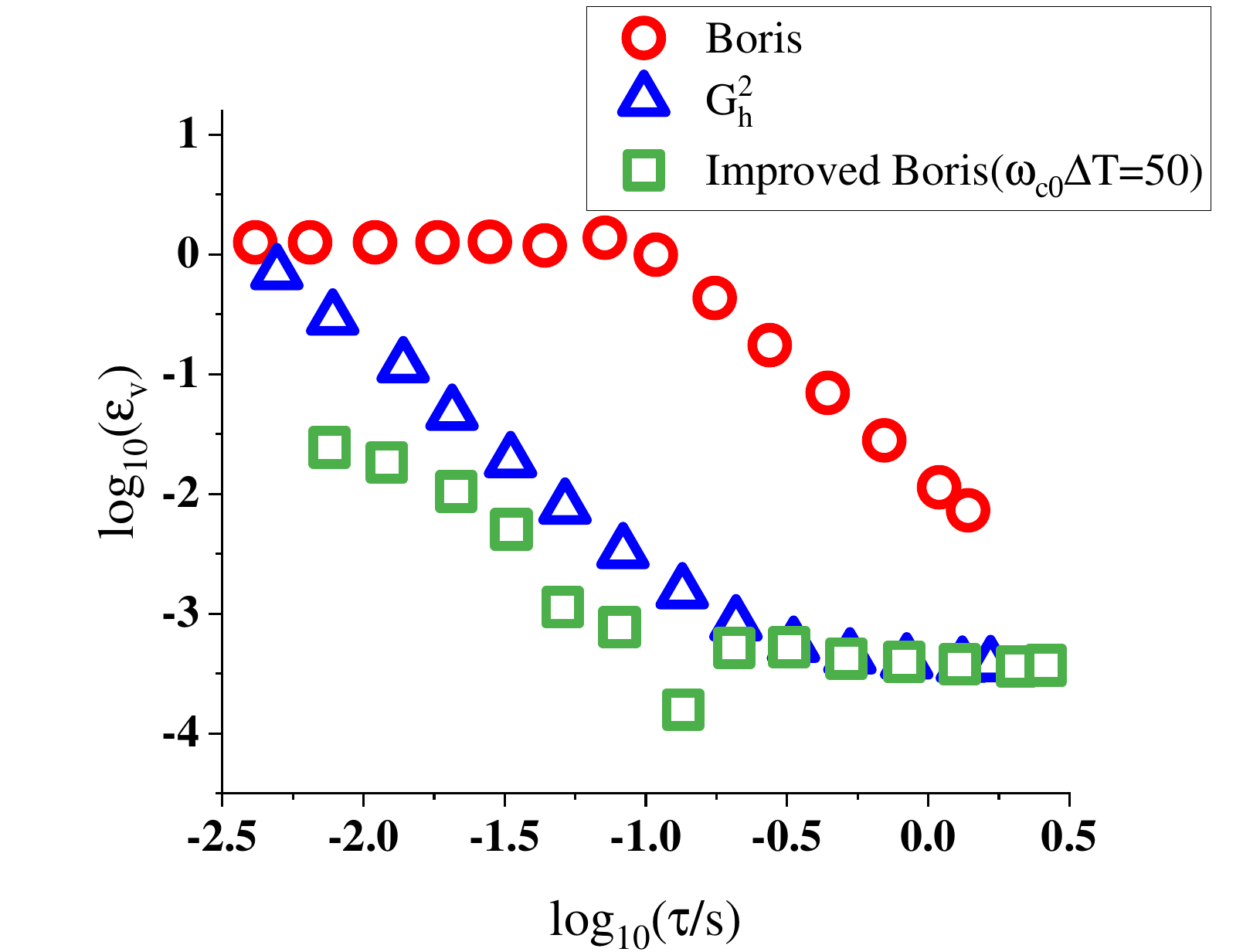}
  \caption{efficiency of velocity}
  \label{fig 5-b}
\end{subfigure}
\caption{Efficiency of all algorithms. Despite the lack of parallelization, the efficiency of the improved Boris algorithm still significantly surpasses that of the other two volume-preserving algorithms. }
\label{fig 5}
\end{figure}

Finally, the efficiency of the algorithms is assessed by considering their accuracy as a function of computational time, as illustrated in Figure \ref{fig 5}. In this figure,data points located closer to the lower-left corner indicate higher efficiency, reflecting both lower error and shorter runtime. Although the Boris algorithm exhibits the shortest computation time, its efficiency is the lowest due to its limited accuracy. In contrast, the efficiency of the improved Boris algorithm markedly surpasses that of the Boris algorithm and $G_h^2$, with this disparity becoming more pronounced under conditions of short computational time (i.e. large time step size). 

To summarize, in the case where only the Tokamak magnetic field is present, the improved Boris algorithm successfully integrates the strengths of both the conventional Boris algorithm and $G_h^2$. Specifically, it preserves phase stability while simultaneously avoiding global trajectory displacement. Despite the absence of parallelization, which leads to a relatively longer computational time under identical conditions, the algorithm's precision and efficiency remain markedly superior to those of the traditional VPAs. In the following subsections, the induced electric field will be incorporated to assess the performance of the new algorithm in addressing wave heating problems.

\subsection{Banana Orbit with high-frequency {\color{blue}full Electromagnetic Field}}

\begin{figure}[tbp]
\centering
\begin{subfigure}[tbp]{0.49\textwidth}
  \includegraphics[width=\textwidth]{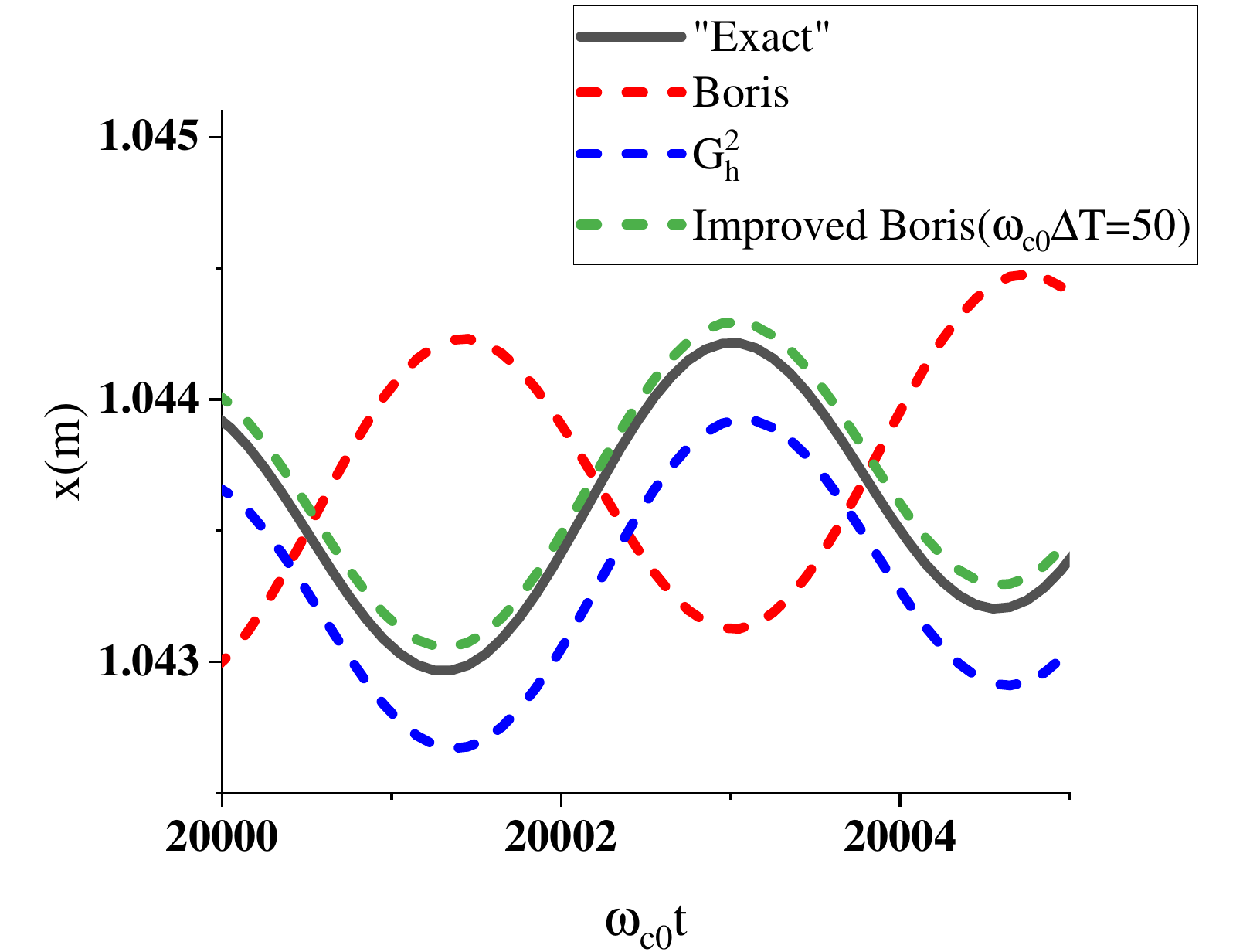}
  \caption{$x$ in [20000,20005]}
  \label{fig 6-a}
\end{subfigure}
\hfill 
\begin{subfigure}[tbp]{0.49\textwidth}
  \includegraphics[width=\textwidth]{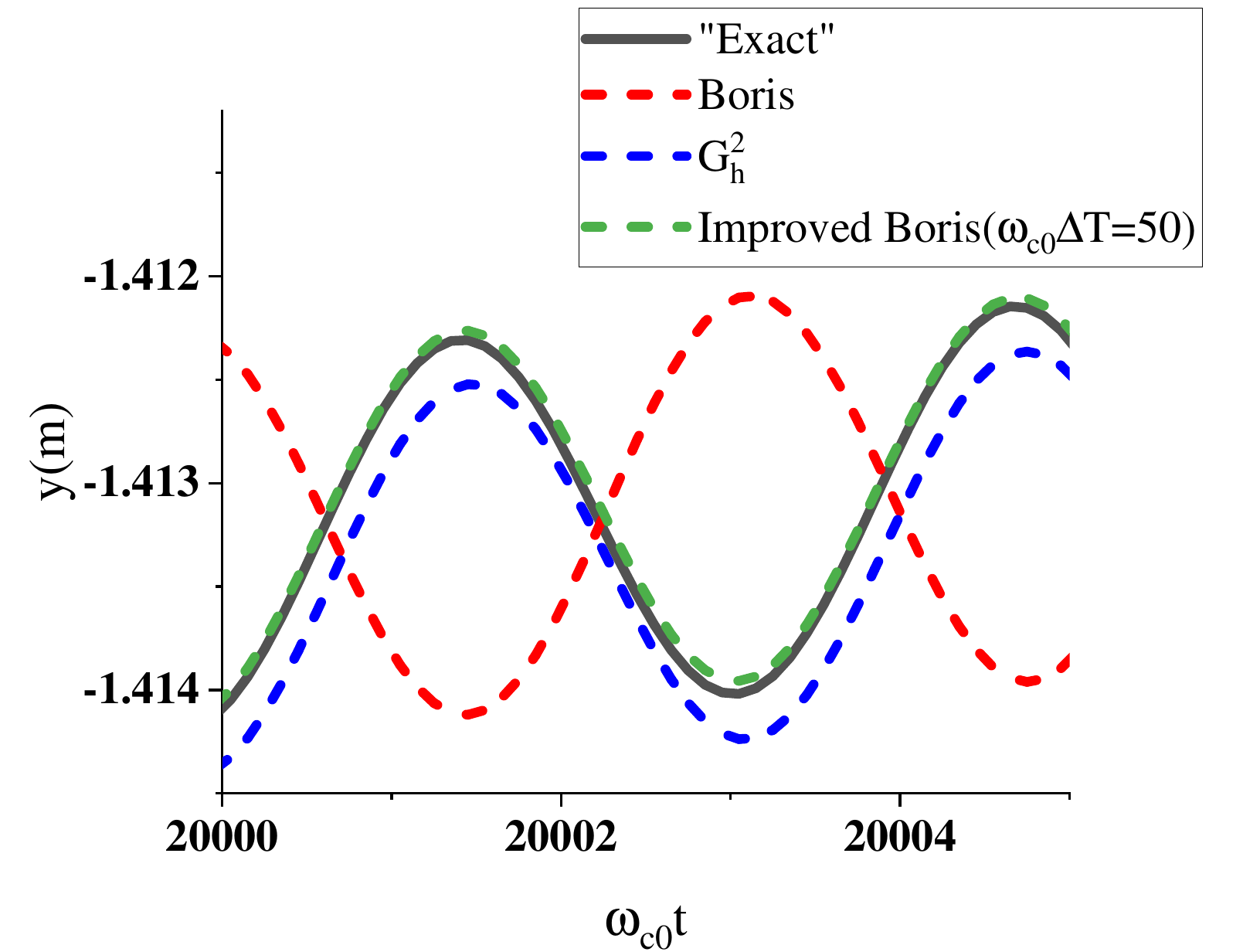}
  \caption{$y$ in [20000,20005]}
  \label{fig 6-b}
\end{subfigure}
\hfill 
\begin{subfigure}[tbp]{0.49\textwidth}
  \includegraphics[width=\textwidth]{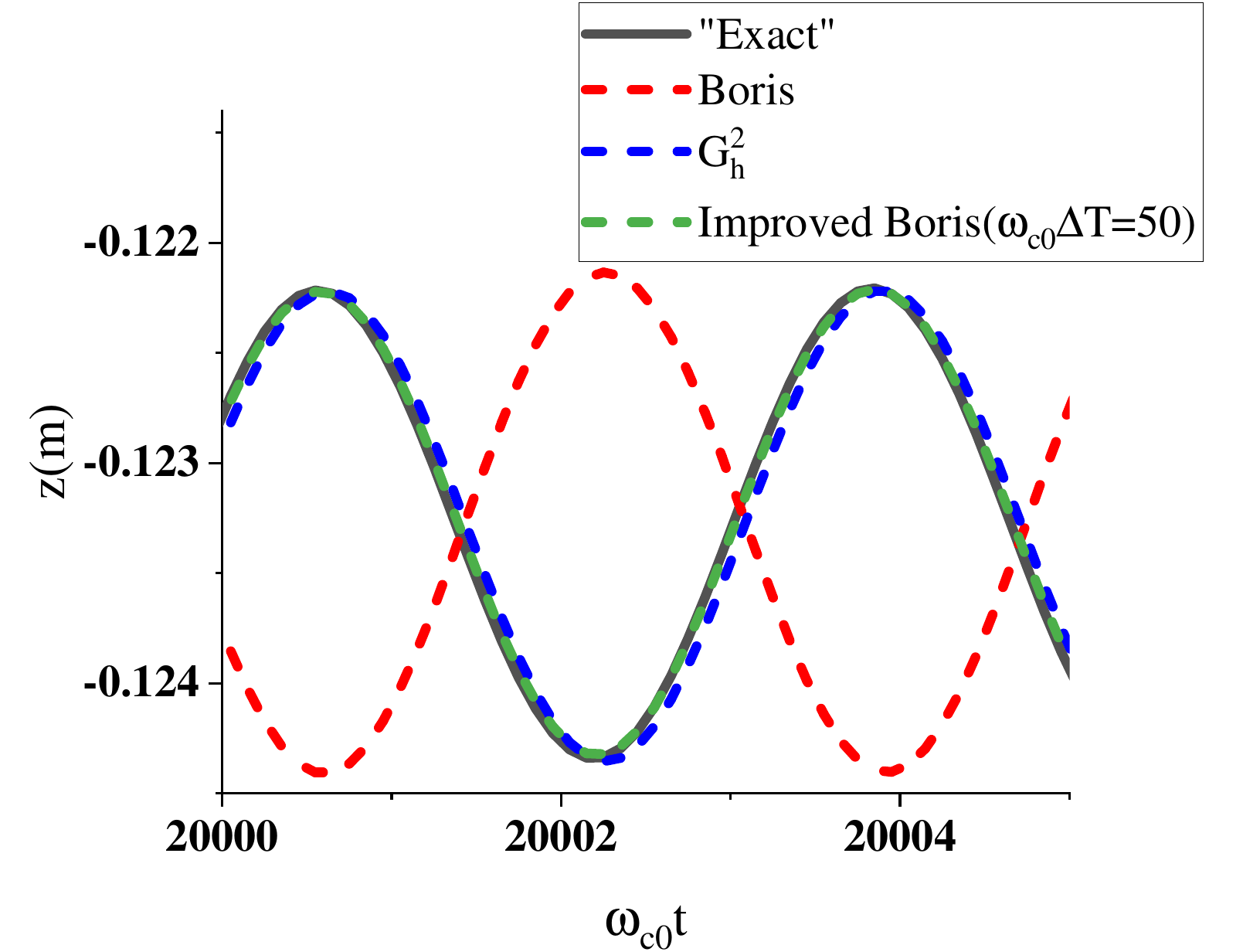}
  \caption{$z$ in [20000,20005]}
  \label{fig 6-c}
\end{subfigure}
\hfill 
\begin{subfigure}[tbp]{0.49\textwidth}
  \includegraphics[width=\textwidth]{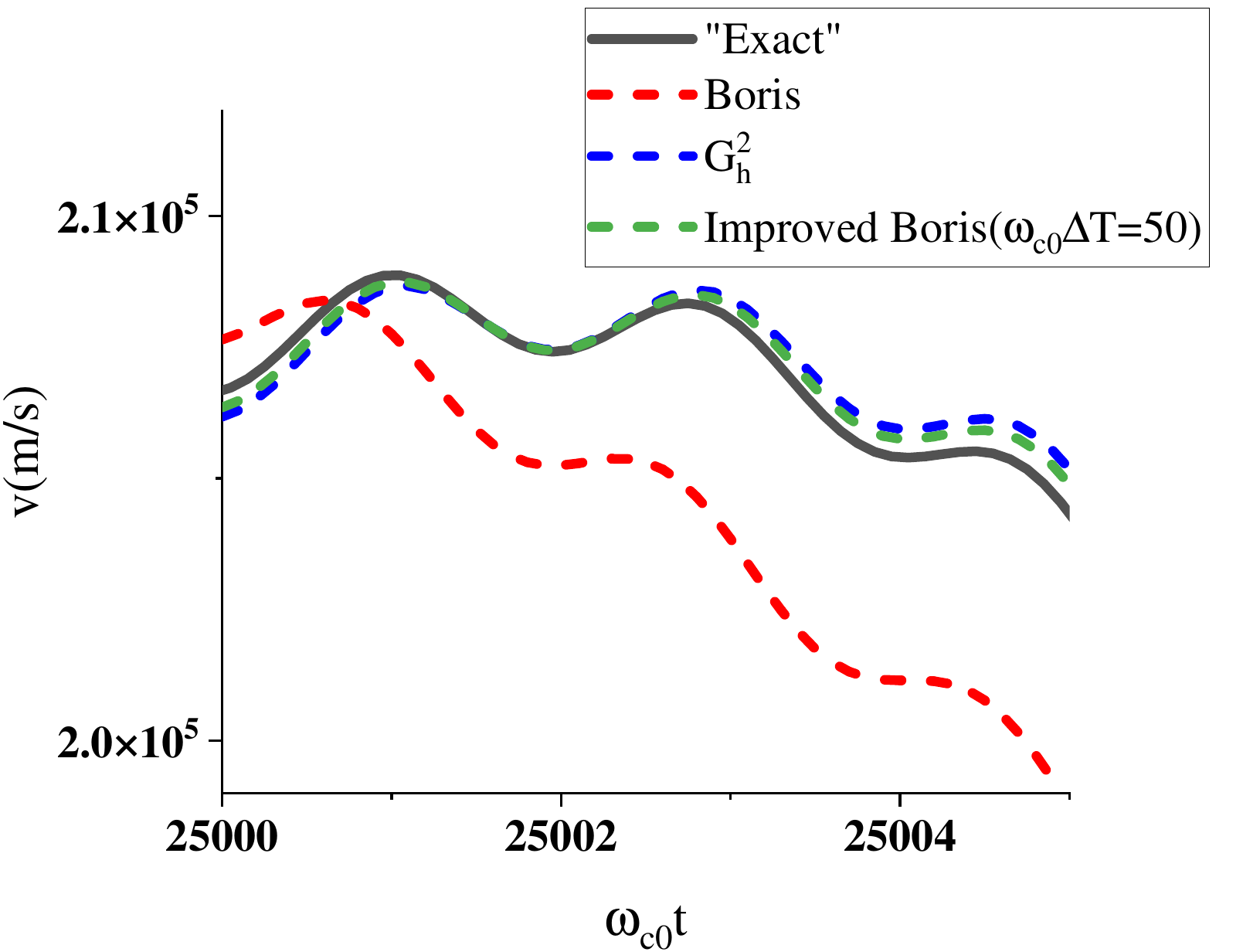}
  \caption{$v$ in [25000,25005]}
  \label{fig 6-d}
\end{subfigure}
\caption{Time-dependent numerical results of the banana orbit with a high-frequency {\color{blue}full Electromagnetic Field} in selected time intervals. Compared to the Boris algorithm {\color{blue}(red dashed lines)} and $G_h^2$ {\color{blue}(blue dashed lines)}, the numerical results obtained by the improved Boris algorithm {\color{blue}(green dashed lines)} remain the closest to the "exact" solution {\color{blue}(black solid lines)}.}
\label{fig 6}
\end{figure}

~~~~{\color{blue}In addition to the background Tokamak magnetic field $\vec B$ given by Equations (15), high-frequency electromagnetic perturbations $\vec B_{1}$ and $\vec E_{1}$ are also introduced in this section
$$
\vec E_{1}(\vec r,t)=(0,0,E_0 \text{cos} (\phi+\omega_0  t))
\eqno{(18.a)}
$$
$$
\vec B_{1}(\vec r,t)=-\frac{1}{\omega_0} \frac{E_0 \text{cos} (\phi+\omega_0  t) }{x^2+y^2}(x,y,0)
\eqno{(18.b)}
$$
Here, $\phi=\text{arctan}(\frac{y}{x})$ denotes the toroidal angle, $E_0=5 \times 10^3\ V/m$, and $\omega_0=1.5\omega_{c0}$ which is of the same order of magnitude as the cyclotron frequency. The above expression characterizes a time-harmonic electromagnetic wave propagating in the toroidal direction with a frequency of $\omega_0$.} And all other computational conditions are kept consistent with those in the previous subsection. Figure \ref{fig 6} presents the time-dependent results for $x,y,z$. And the magnitude of velocity, $v=|\vec v|$, instead of its components, is displayed to provide a clearer evaluation of the effectiveness of wave heating. Compared to the case without electric fields, the numerical solution obtained by the improved Boris algorithm is no longer visually indistinguishable from the “exact” solution. Nevertheless, it continues to show the closest agreement among all three methods. Both the cyclotron phase and guiding center motion are accurately captured. 

\begin{figure}[tbp]
\centering
\begin{subfigure}[tbp]{0.49\textwidth}
  \includegraphics[width=\textwidth]{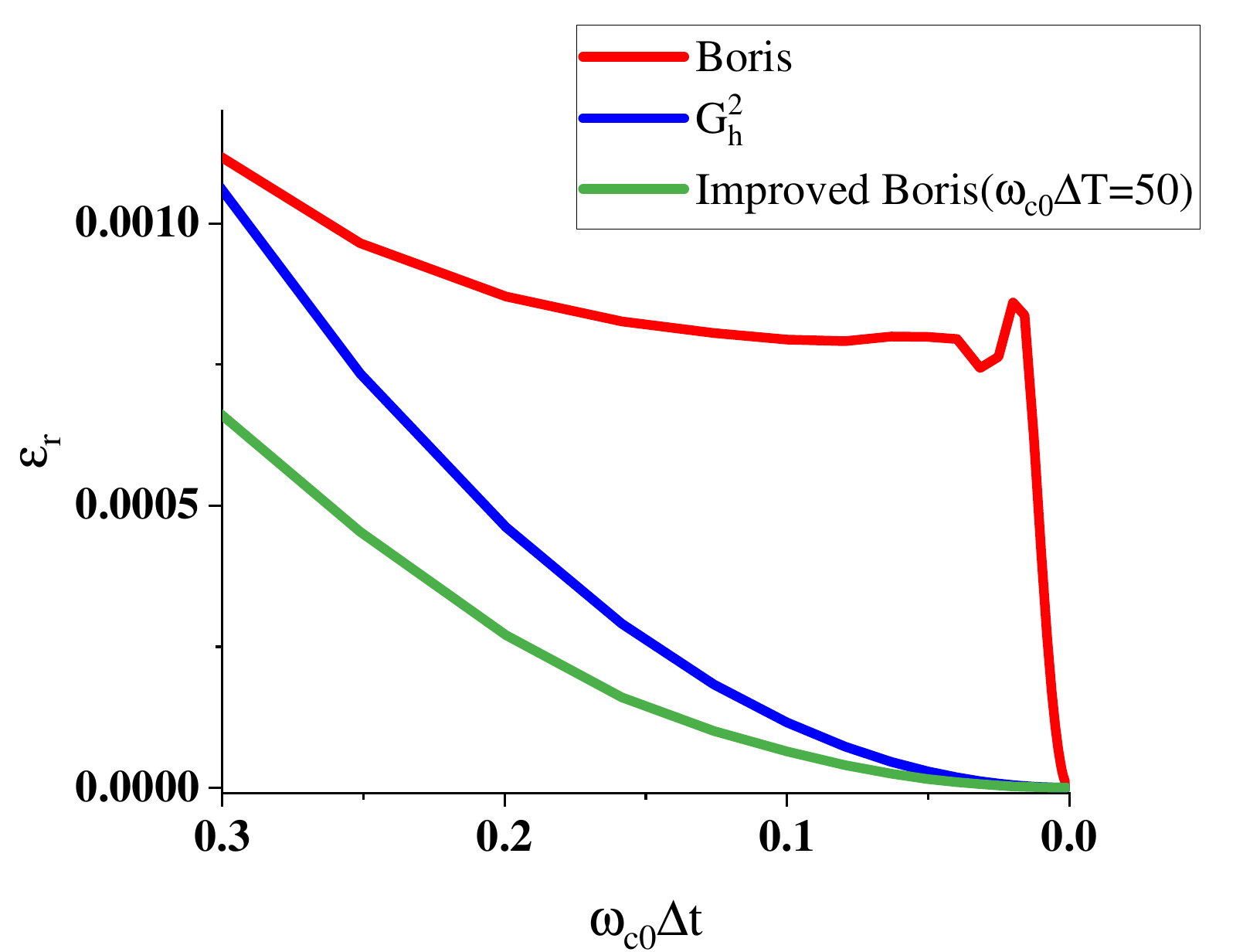}
  \caption{average relative errors of position}
  \label{7-a}
\end{subfigure}
\hfill 
\begin{subfigure}[tbp]{0.49\textwidth}
  \includegraphics[width=\textwidth]{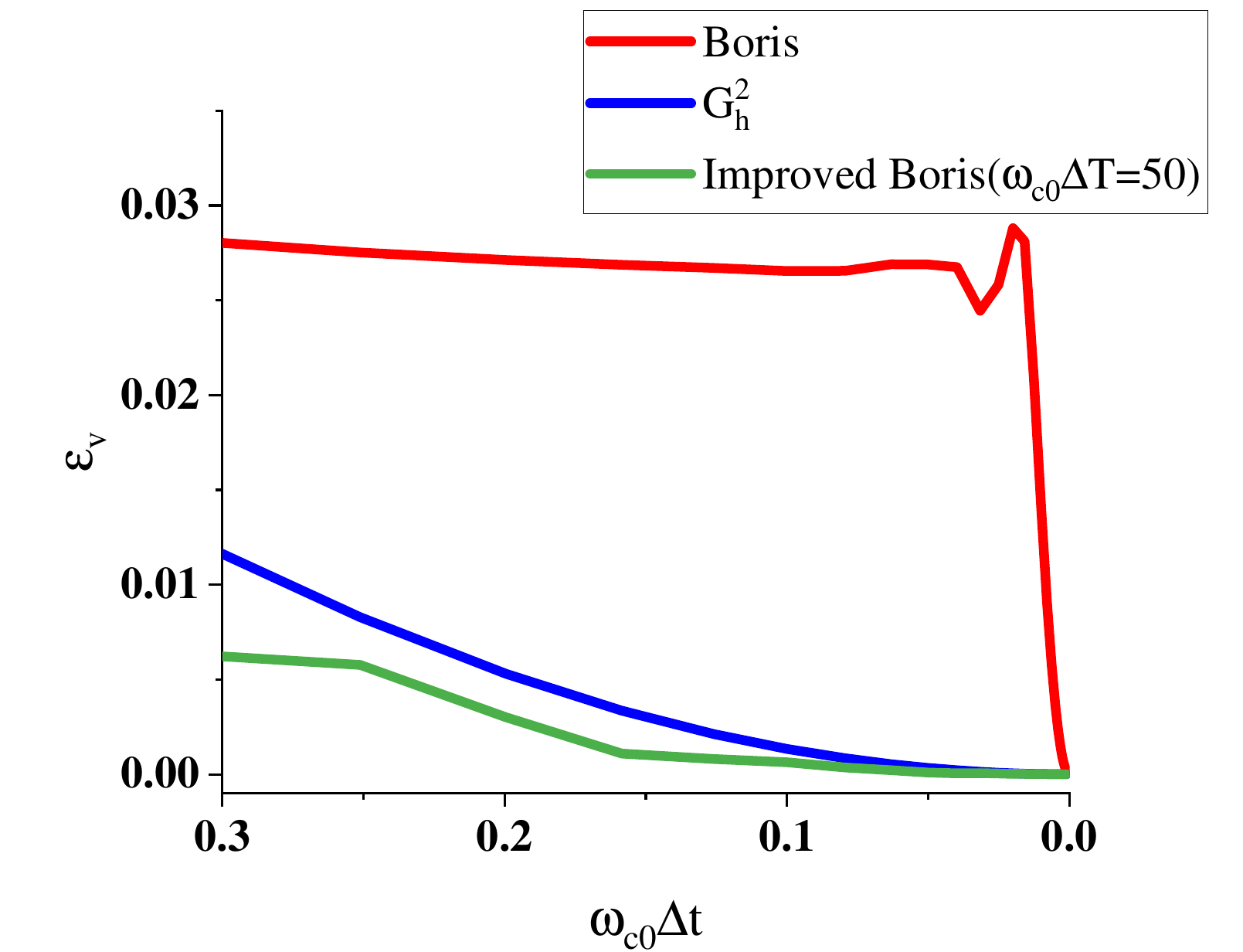}
  \caption{average relative errors of velocity magnitude}
  \label{fig 7-b}
\end{subfigure}
\caption{Average relative errors of $\vec r$ and $v$ as functions of time step size $\Delta t$ by all algorithms. Under the conditions of this subsection, the advantages of the improved Boris algorithm are less pronounced; however, it continues to demonstrate a distinct superiority.}
\label{fig 7}
\end{figure}

The average relative error of position $\epsilon_{\vec r}$ is calculated by Equation (17.a) and presented as a function of time step size $\Delta t$ in Figure \ref{fig 7}, as well as the average relative error of velocity magnitude $\epsilon_v$ given by
$$
\epsilon_v = \frac{1}{N} \sum_{m=0}^{N-1} \frac{|v_m^{"exact"}-v_m^{"numerical"}|}{|v_m^{"exact"}|}
\eqno{(19)}
$$
Unlike the previous subsection, logarithmic scaling is not applied in this figure. In this particular case, where the electric field frequency $\omega_0$ is comparable to the gyro-frequency, $G_h^2$, which provides a more accurate resolution of the cyclotron motion, shows a clear performance advantage over the conventional Boris algorithm. This aligns with expectations based on its design for high-frequency dynamics. Although the improved Boris algorithm does not demonstrate the same order-of-magnitude superiority observed in the absence of electric fields, it still consistently outperforms $G_h^2$ in most scenarios—achieving over 50\% reduction in average error across a wide range of time step sizes. 

\subsection{Transit Orbit with low-frequency Electric Field}
\begin{figure}[tbp]
\centering
\includegraphics[width=0.6\textwidth]{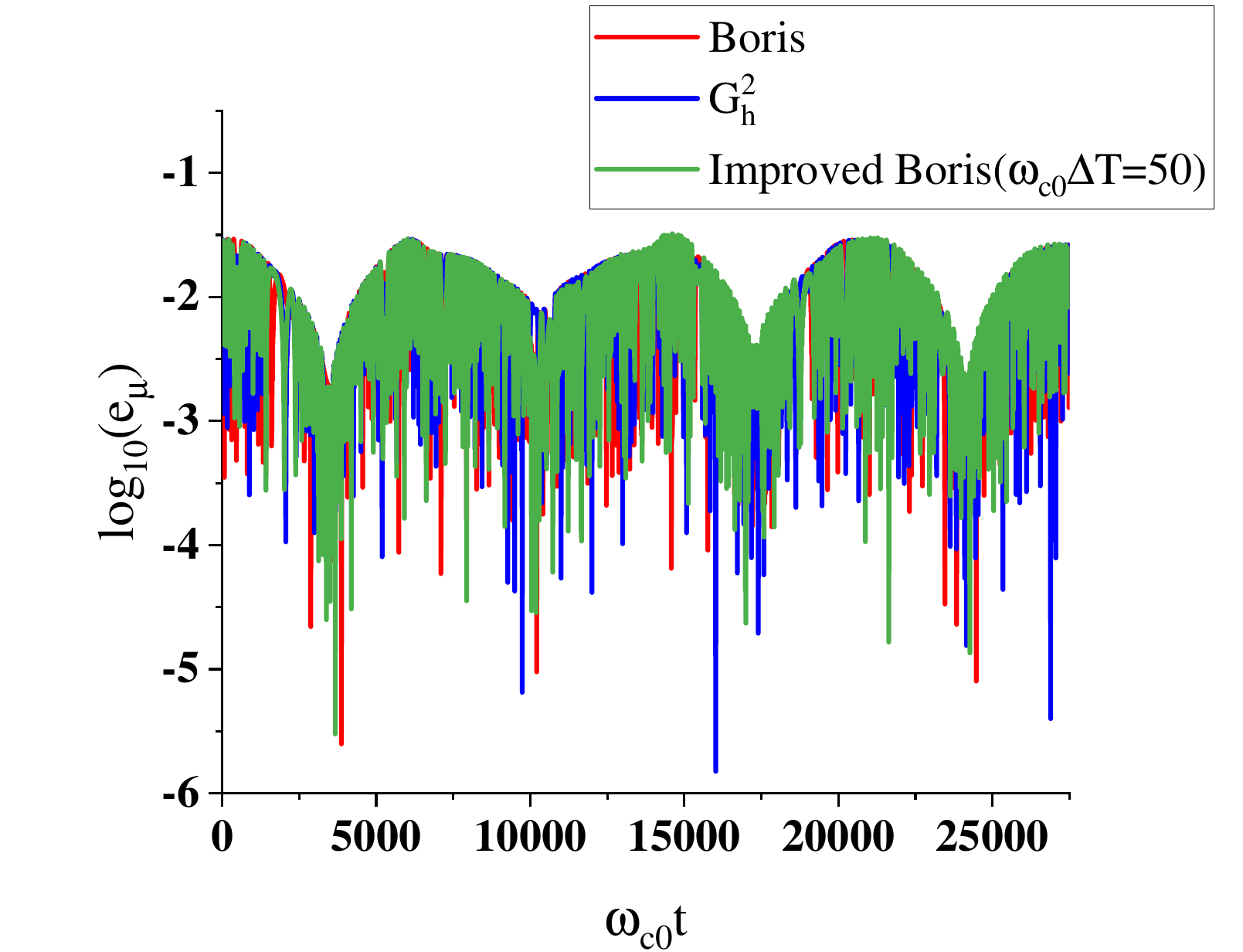}
\caption{Time-dependent relative errors of magnetic moment in $[0,2T_1]$ by all algorithms.}
\label{fig 8-1}
\end{figure}

~~~~In this subsection, the initial velocity is modified to {\color{blue}$\vec v_0=(0,8\times 10^4\ m/s,2\times 10^5\ m/s)$}. This alteration results in a shift in the particle's trajectory from banana orbit to transit orbit, and the transit period $T_1$ is approximately $\omega_{c0} T_1=1.38 \times 10^4$. Meanwhile, a low-frequency electric field at the transit frequency, i.e. $\vec E=(0,0,E_0{\color{blue}\text{cos}}(\omega_1 t))$, {\color{blue}$E_0=5 \times 10^3\ V/m$} with $\omega_1=\frac{2\pi}{T_1}$, is also considered. The time integration interval is set to $[0,2T_1]$. {\color{blue}In this scenario, as the external fields evolve slowly compared to the gyro-period, the magnetic moment $\mu$ is expected to remain an adiabatic invariant. The time evolution of the relative error in $\mu$ for all three algorithms is shown in Figure \ref{fig 8-1}, confirming that all methods effectively preserve this invariant throughout the simulation.}

\begin{figure}[tbp]
\centering
\begin{subfigure}[tbp]{0.49\textwidth}
  \includegraphics[width=\textwidth]{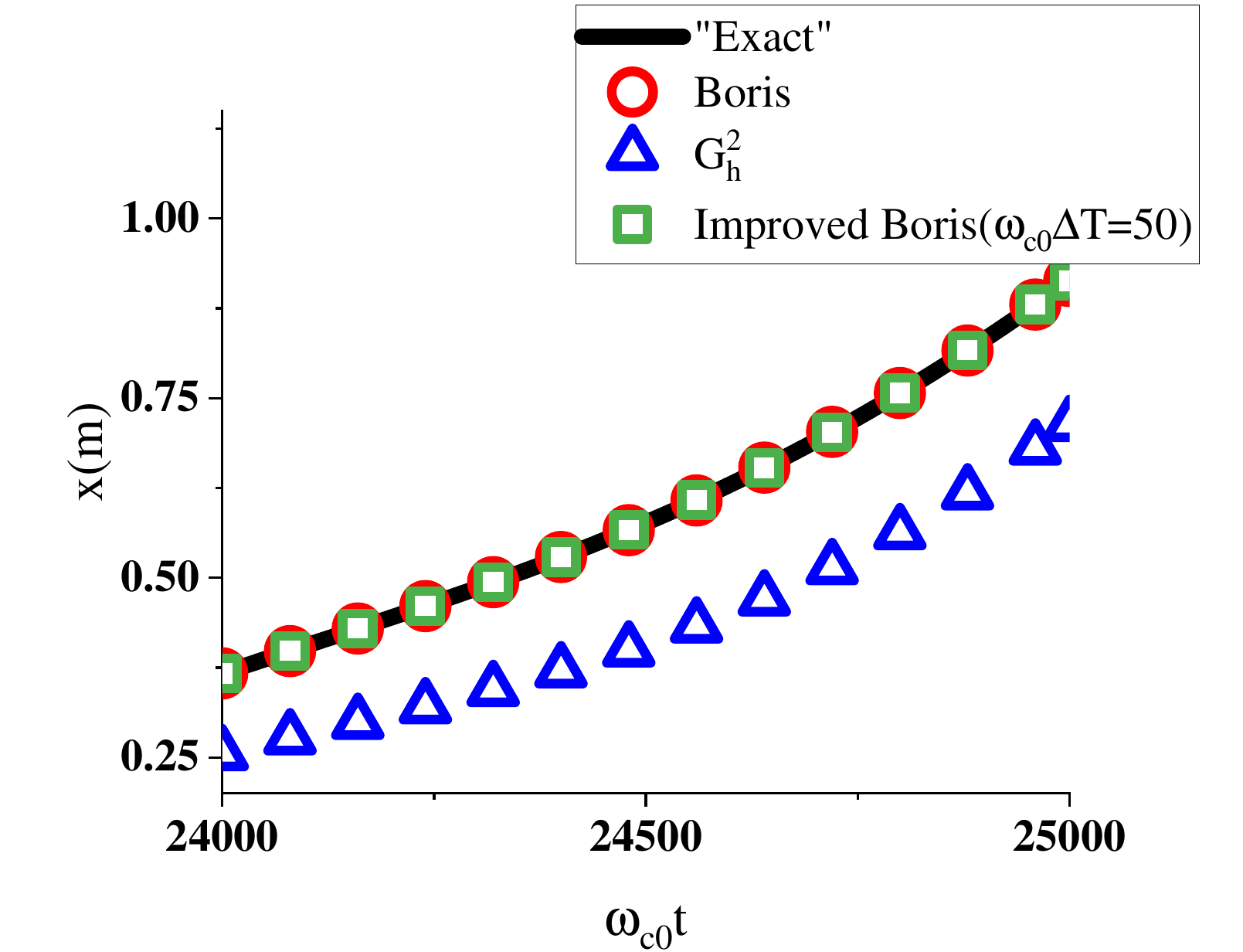}
  \caption{$x$ in [24000,25000]}
  \label{fig 8-a}
\end{subfigure}
\hfill 
\begin{subfigure}[tbp]{0.49\textwidth}
  \includegraphics[width=\textwidth]{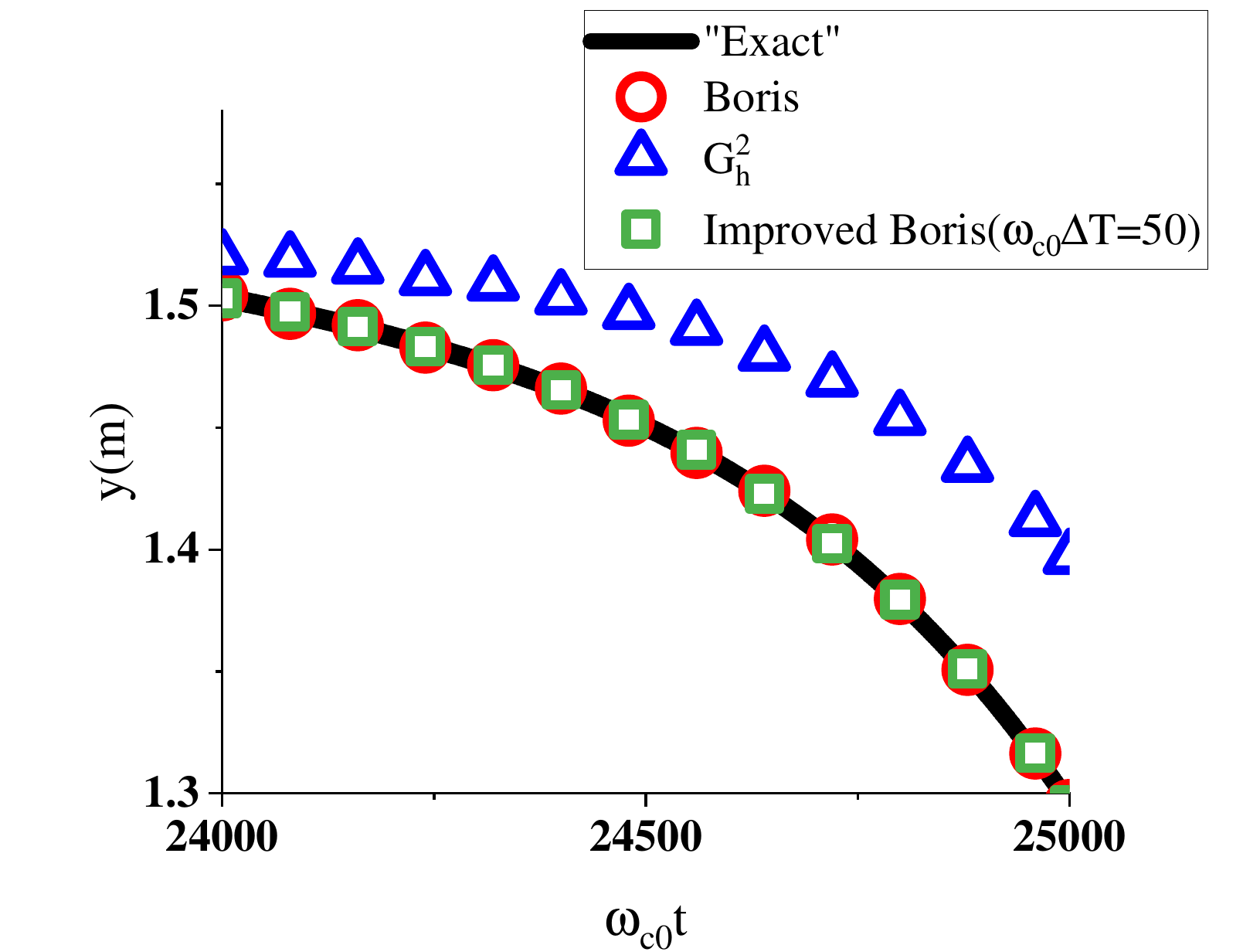}
  \caption{$y$ in [24000,25000]}
  \label{fig 8-b}
\end{subfigure}
\hfill 
\begin{subfigure}[tbp]{0.49\textwidth}
  \includegraphics[width=\textwidth]{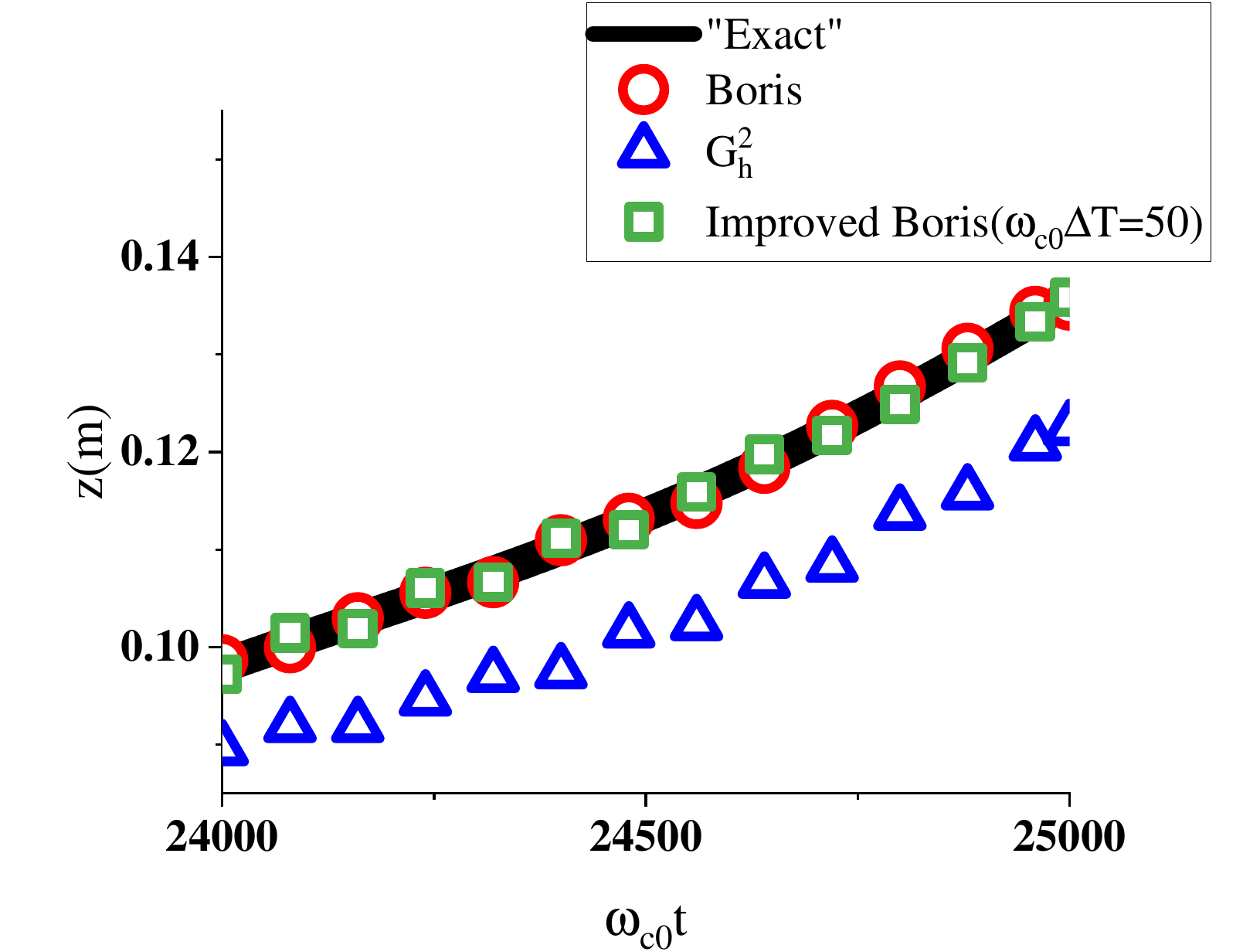}
  \caption{$z$ in [24000,25000]}
  \label{fig 8-c}
\end{subfigure}
\hfill 
\begin{subfigure}[tbp]{0.49\textwidth}
  \includegraphics[width=\textwidth]{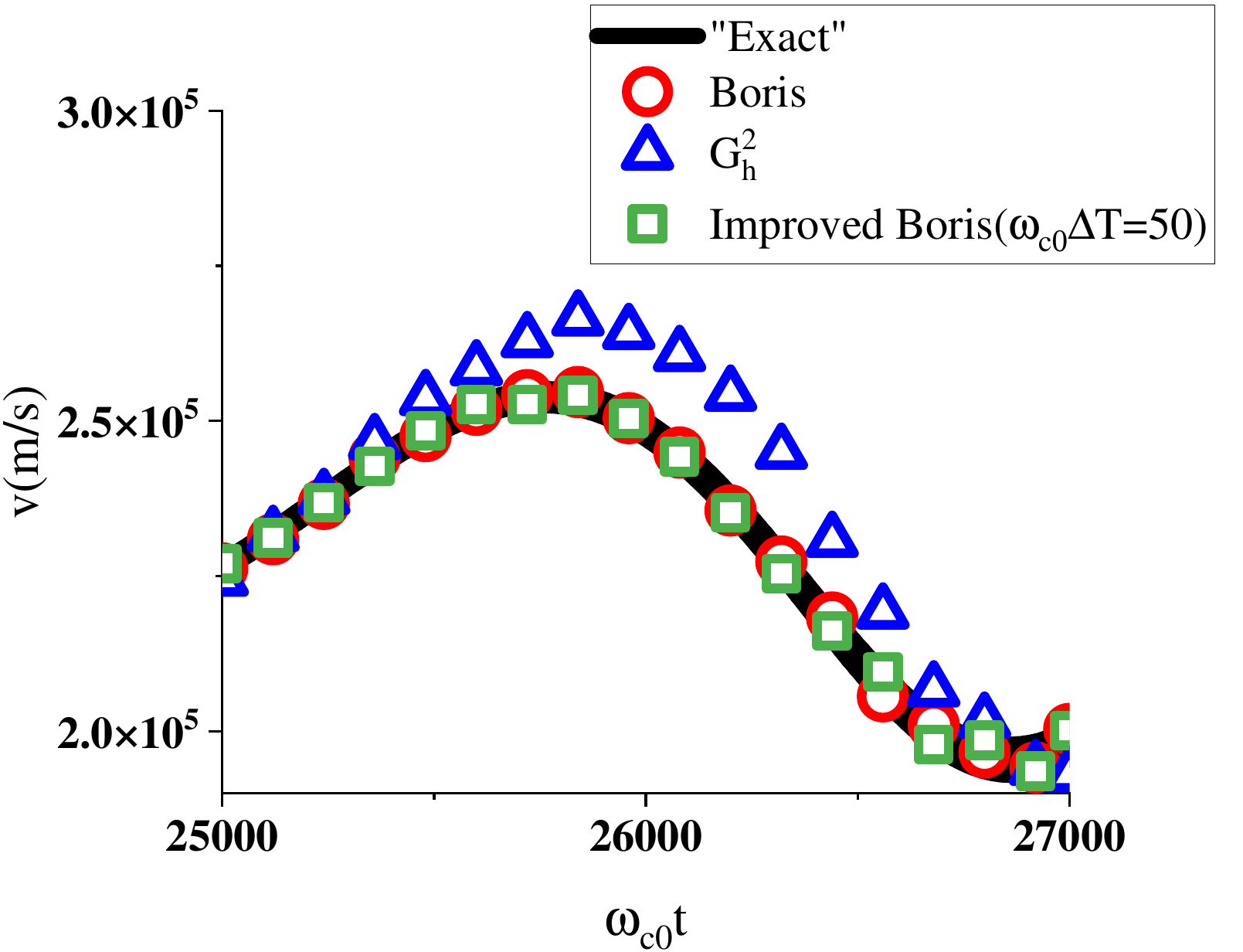}
  \caption{$v$ in [25000,27000]}
  \label{fig 8-d}
\end{subfigure}
\caption{Time-dependent numerical results of the transit orbit with a low-frequency resonant electric field in relative long selected time intervals. Compared to the other two algorithms, the numerical results by $G_h^2$ exhibit a noticeable deviation.}
\label{fig 8}
\end{figure}

In Figure \ref{fig 8}, we present the numerical solutions for $\vec r$ and $v$ over extended time intervals. The three components of $\vec{r}$ are shown in sub-figures (a), (b), and (c) over the interval $[24000,25000]$. Even on this longer timescale, the trajectory computed using $G_h^2$ exhibits noticeable deviation from the "exact" solution. This discrepancy arises due to the algorithm’s inherent limitations in handling low-frequency dynamics, which are further exacerbated by the low-frequency resonant electric field. Meanwhile, the results of the other two algorithms essentially coincide with the "exact" solution, with no visible deviation, demonstrating their robustness under slow field variation. As depicted in sub-figure (d), the velocity magnitude $v$ within $[25000,27000]$ further indicates that $G_h^2$ fails to accurately capture the wave-particle interaction under low-frequency excitation, while both the Boris algorithm and the improved Boris algorithm continue to perform reliably. 

\begin{figure}[tbp]
\centering
\begin{subfigure}[tbp]{0.49\textwidth}
  \includegraphics[width=\textwidth]{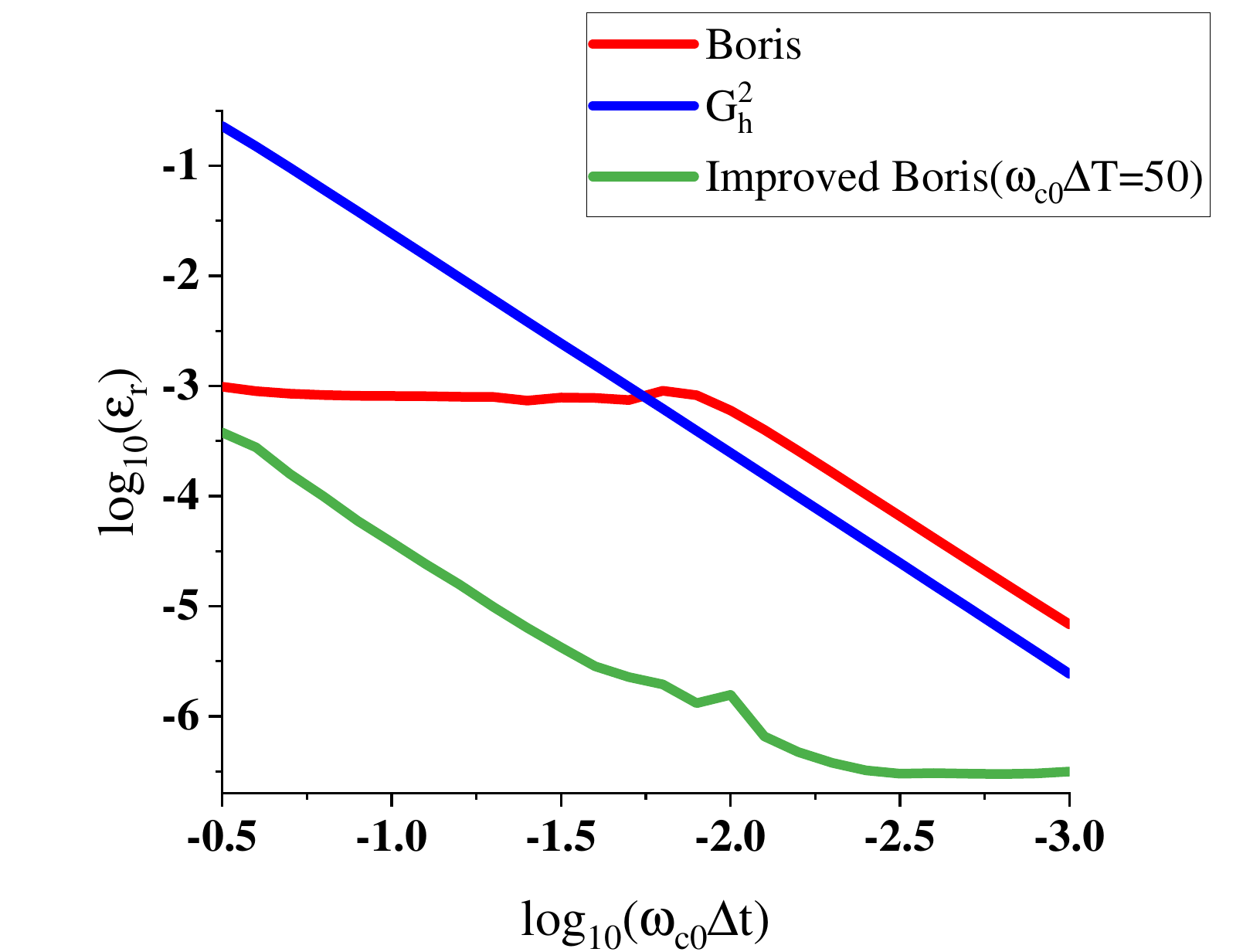}
  \caption{average relative errors of position}
  \label{fig 9-a}
\end{subfigure}
\hfill 
\begin{subfigure}[tbp]{0.49\textwidth}
  \includegraphics[width=\textwidth]{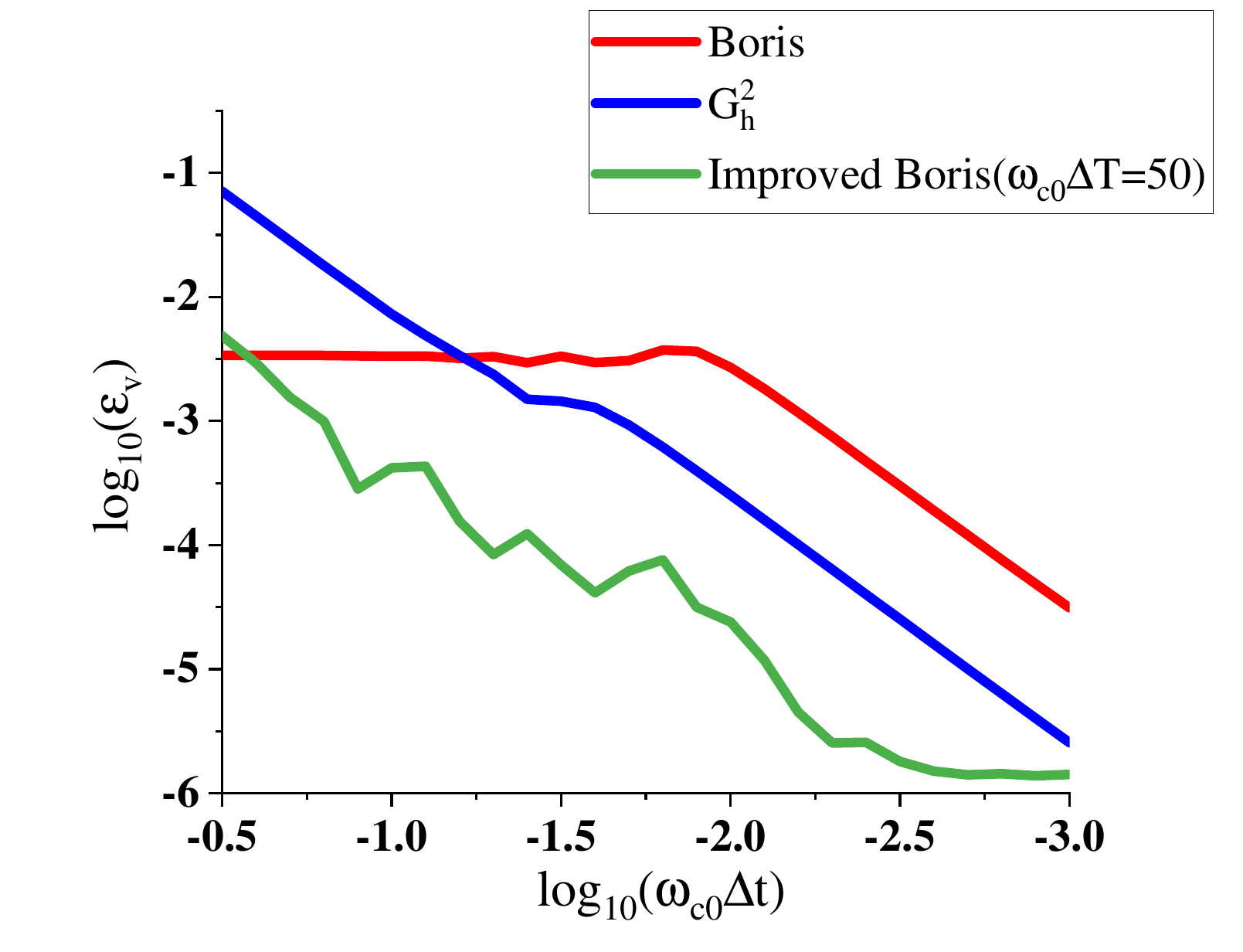}
  \caption{average relative errors of velocity magnitude}
  \label{fig 9-b}
\end{subfigure}
\caption{Average relative errors of $\vec r$ and $v$ as functions of time step size $\Delta t$ by all algorithms. Under the conditions of this subsection, the advantages of the improved Boris algorithm even surpass those observed in the case without the magnetic field.}
\label{fig 9}
\end{figure}

The convergence curves of the average relative errors, as shown in Figure \ref{fig 9} {\color{blue}(with logarithmic scaling)}, reveal that the Boris algorithm exhibits a distinct advantage over $G_h^2$ at larger time step sizes. However, due to the near-constant accuracy of the Boris algorithm for time steps larger than $10^{-2} \omega_{c0}^{-1}$, the precision of $G_h^2$ eventually surpasses that of the Boris algorithm at smaller time step sizes. This behavior is consistent with the properties of the Boris algorithm: its cyclotron phase convergence begins only when the time step falls below a critical threshold, which in this case appears near $10^{-2} \omega_{c0}^{-1}$. The improved Boris algorithm, however, maintains a substantial advantage. In most cases, its accuracy exceeds that of the better-performing VPA by one to two orders of magnitude.

\subsection{Banana Orbit with Electric Field of Various Frequencies}
~~~~To more precisely evaluate the performance of the three methods in wave heating problems across various frequencies, this subsection adopts the initial conditions of the banana orbit, with the time integration interval adjusted to $[0,T_0]$, a fixed time step size of $\omega_{c0} \Delta t=0.1$, and a resonant electric field along the z-axis given by $\vec E=(0,0,E_0{\color{blue}\text{cos}}(\omega t))$, {\color{blue}$E_0=5 \times 10^3\ V/m$}. The average relative errors of position and velocity magnitude are treated as functions of the frequency of the electric field $\omega$. 

\begin{figure}[tbp]
\centering
\begin{subfigure}[tbp]{0.49\textwidth}
  \includegraphics[width=\textwidth]{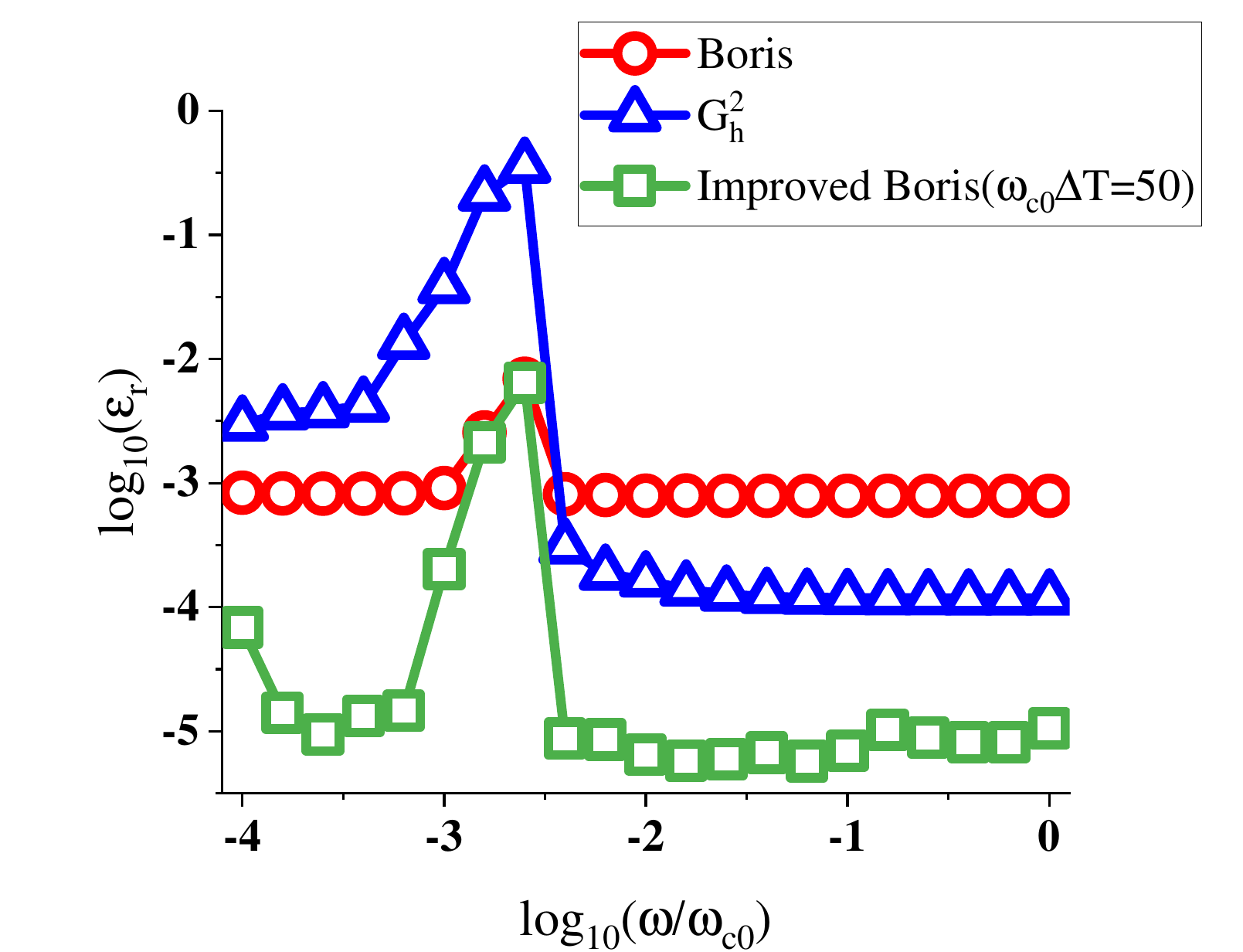}
  \caption{average relative errors of position at low characteristic frequencies}
  \label{fig 10-a}
\end{subfigure}
\hfill 
\begin{subfigure}[tbp]{0.49\textwidth}
  \includegraphics[width=\textwidth]{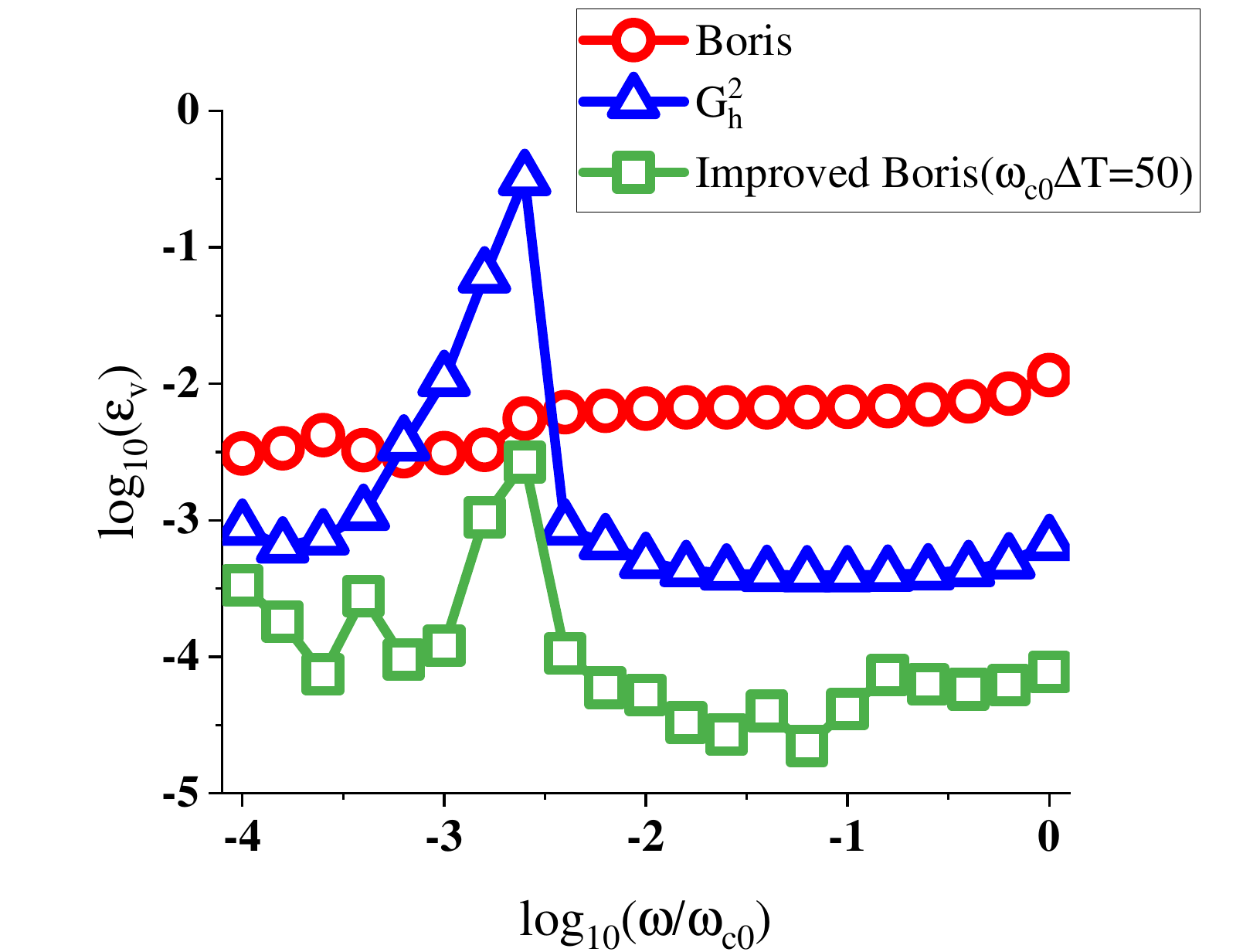}
  \caption{average relative errors of velocity magnitude at low characteristic frequencies}
  \label{fig 10-b}
\end{subfigure}
\caption{Average relative errors of $\vec r$ and $v$ as functions of the frequency of the electric field by all algorithms, low-frequency cases.}
\label{fig.10}
\end{figure}

In Figure \ref{fig.10}, the low-frequency cases are illustrated, where the value of $\omega$ ranges from $10^{-4}\omega_{c0}$ to $\omega_{c0}$. The results are plotted on logarithmic scaling for improved clarity. The accuracy of the Boris algorithm remains virtually constant across the entire frequency range. In contrast, $G_h^2$ exhibits a non-monotonic trend: its accuracy initially degrades with increasing frequency, reaching a minimum, then begins to improve, eventually surpassing the Boris algorithm near $\omega / \omega_{c0} \approx 10^{-2.5}$. It then stabilizes at a level approximately one order of magnitude more accurate than the Boris algorithm. The improved Boris algorithm consistently delivers the best performance. Even in the least favorable cases, its accuracy is comparable to that of the Boris algorithm, while in most cases, it exceeds the better-performing VPA by one to two orders of magnitude. It is evident that the improved Boris algorithm constitutes a more dependable option for addressing low-frequency problems.

\begin{figure}[tbp]
\centering
\begin{subfigure}[tbp]{0.49\textwidth}
  \includegraphics[width=\textwidth]{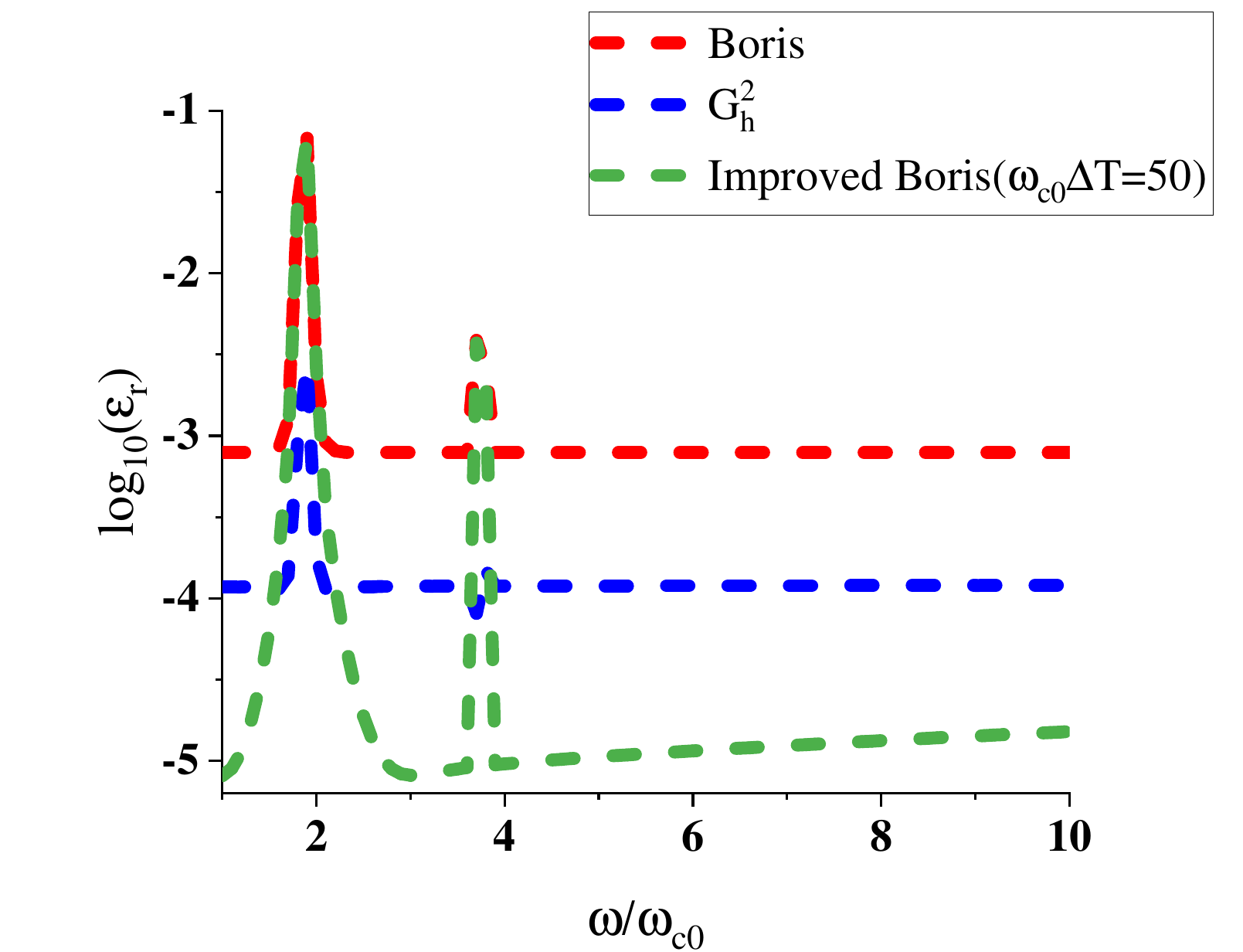}
  \caption{average relative errors of position at high characteristic frequencies}
  \label{fig 11-a}
\end{subfigure}
\hfill 
\begin{subfigure}[tbp]{0.49\textwidth}
  \includegraphics[width=\textwidth]{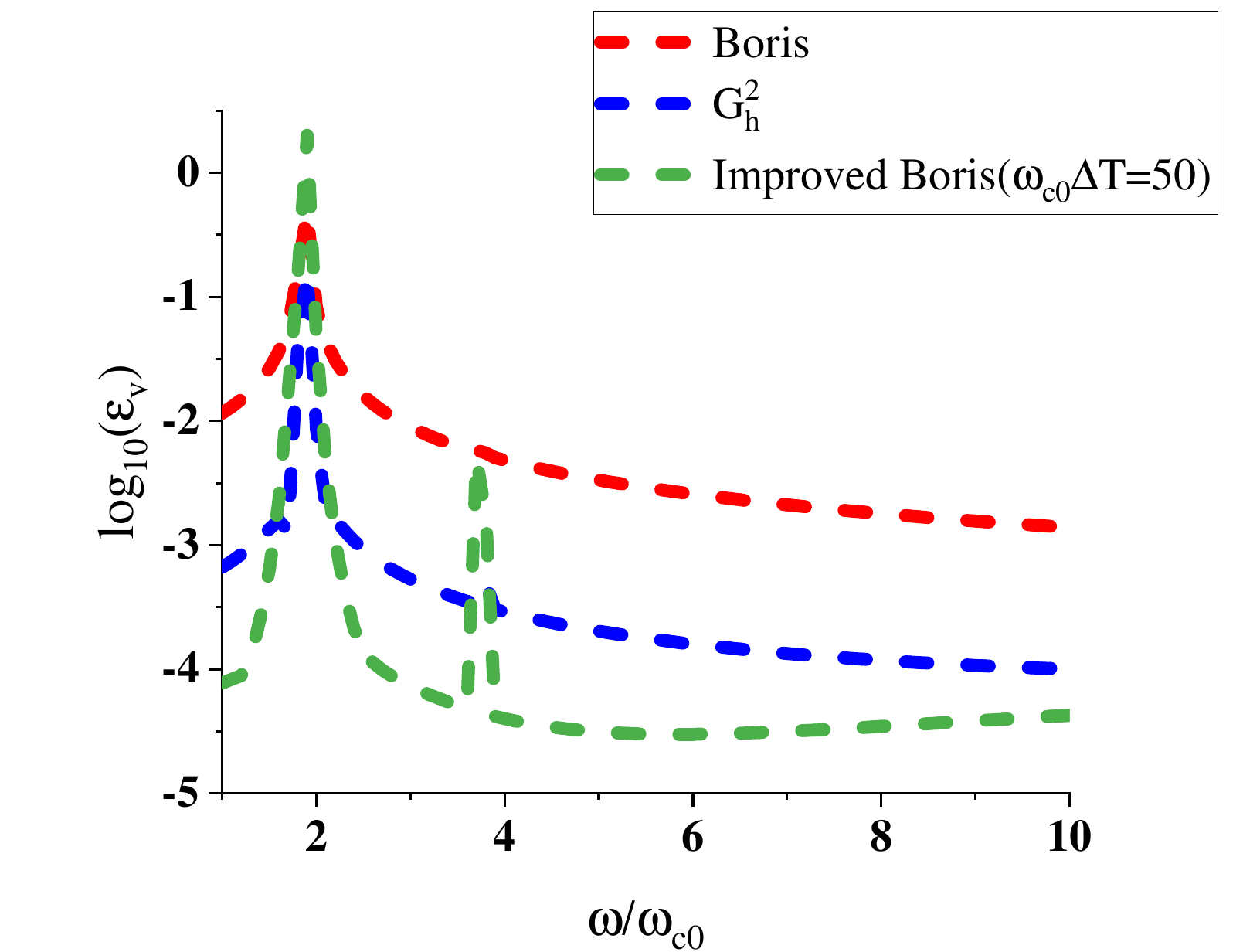}
  \caption{average relative errors of velocity magnitude at high characteristic frequencies}
  \label{fig 11-b}
\end{subfigure}
\caption{Average relative errors of $\vec r$ and $v$ as functions of the frequency of the electric field by all algorithms, high-frequency cases. The two unstable regions depicted in this figure correspond to the particle's cyclotron frequency $\omega_c$ and $2\omega_c$, respectively.}
\label{fig.11}
\end{figure}

Figure~\ref{fig.11} illustrates the results for relatively high-frequency electric fields, where $\omega \in [\omega_{c0},10\omega_{c0}]$. The accuracy trends of the three algorithms exhibit similar behavior across this frequency range: all methods display localized instability near $\omega \approx 2\omega_{c0}$ and $\omega \approx 4\omega_{c0}$, while maintaining relatively stable accuracy elsewhere. Within the stable region, the improved Boris algorithm maintains the highest accuracy, followed by $G_h^2$, and then the Boris algorithm, with each adjacent pair differing by approximately one order of magnitude. However, within the unstable regions, the accuracy of the improved Boris algorithm degrades significantly, becoming comparable to that of the Boris algorithm.      

{\color{blue}For the Tokamak magnetic field configuration considered in this paper, the cyclotron frequency on the magnetic axis is $2\omega_{c0}$. And the initial conditions correspond to a deeply trapped particle located very close to the magnetic axis, implying that its local cyclotron frequency $\omega_c$ remains around $2\omega_{c0}$ throughout its motion. Consequently, it can be inferred that the unstable regions depicted in the figure are located in the vicinity of $\omega_c$ and $2\omega_c$.  

In the case of $\omega \approx \omega_c$, the system enters a strong resonance regime, where the proximity of the electric field frequency to the cyclotron frequency results in substantial energy absorption. This leads to a rapid increase in particle kinetic energy and causes severe trajectory instability. Even under the relatively modest electric field amplitude used in this test, the particle velocity increases by nearly 20-fold after one banana period $T_0$. In this scenario, even the most accurate method—$G_h^2$—exhibits velocity errors exceeding 10\%, indicating that none of the tested VPAs, including the improved Boris algorithm, are effective under such strong resonance conditions. This performance breakdown can be attributed to a fundamental limitation in the algorithmic design: all methods assume a constant electric field within each time step, an assumption that becomes invalid under strong field-particle coupling at resonance. Properly capturing such interactions requires incorporating the temporal variation of the electric field into the single-step update, which will be the focus of our future work. The case of $\omega \approx 2\omega_c$ exhibits a similar but less pronounced instability.} 

In general, for problems at various characteristic frequencies, as long as the traditional VPAs {\color{blue}(especially the conventional Boris algorithm)} remain valid, the improved Boris algorithm will invariably possess superior accuracy and efficiency. 

\section{Conclusions}
~~~~In this paper, we proposed an improved Boris algorithm that integrates the respective advantages of two representative second-order volume-preserving algorithms (VPAs): the well-known conventional Boris algorithm, which excels at resolving low-frequency guiding center dynamics, and $G_h^2$, which is optimized for high-frequency cyclotron motion. The improved algorithm is also explicitly designed to support parallel implementation. Through comphrehensive test particle simulations in a typical Tokamak magnetic field, the performance of the improved Boris algorithm is compared in detail with traditional VPAs. Numerical results indicates that it effectively overcomes the limitation of the second-order VPAs which can only accurately capture either low- or high-frequency dynamics, and demonstrates superior accuracy and efficiency across a broad spectrum of characteristic frequencies. It holds an absolute advantage in the low-frequency scenario. As for the high-frequency cases, despite a decline in accuracy near $\omega_c$ and $2\omega_c$—regions where traditional VPAs also fail—the improved algorithm maintains stability and reliable performance elsewhere. It is anticipated the improved Boris algorithm holds significant potential for enhancing computational efficiency in large-scale, long-duration plasma simulations.

Still, the limitations near the cyclotron frequency, shared by both the improved Boris algorithm and traditional VPAs, continue to restrict their utility in handling problems like full orbit simulations of ion cyclotron resonance heating {\color{blue}(ICRH)}. A potential solution is to couple the electric and magnetic fields during the single-step time advancement, instead of treating them as separate variables. This extension will form a key part of our future work. {\color{blue} Besides, we will also attempt to extend the proposed construction to relativistic regimes and to higher-order volume-preserving algorithms, as well as to utilize the paralellized plan outlined in Figure \ref{fig 1-1} with actual parallel computing platforms.}

\section*{Acknowledgments}
~~~~This work was supported by the National Natural Science Foundation of China under Grant No. 12205339.

\nocite{*}

\printbibliography
\end{document}